\documentclass[a4paper,12pt]{article}
\pdfoutput=1 

\usepackage{amsmath,amssymb}
\usepackage[pdftex]{hyperref}
\hypersetup{colorlinks=true,linkcolor=blue,urlcolor=blue,filecolor=blue,citecolor=blue}
\usepackage{epsf}
\usepackage{cite}
\usepackage{fancyhdr}
\usepackage{enumerate} 
\usepackage[usenames]{xcolor}

\usepackage{pgfplots}
\usetikzlibrary{decorations.markings}

\makeatletter

\newcommand{\av}{{\bf a}}
\newcommand{\fv}{{\bf f}}
\newcommand{\xv}{{\bf x}}
\newcommand{\yv}{{\bf y}}

\newcommand{\ALM}{{\bf A}}
\newcommand{\BLM}{{\bf B}}
\newcommand{\betaLM}{{\boldsymbol{\beta}}}
\newcommand{\omegaLM}{{\boldsymbol{\omega}}}

\newcommand{\Lap}{\bigtriangleup}

\g@addto@macro\bfseries{\boldmath}

\def\refchecklabelfontsize{\fontsize{5pt}{5pt}\selectfont}
\let\mark@size=\refchecklabelfontsize

\def\half{{\frac{1}{2}}}

\def\p{\partial}

\def\unit{{1\kern-.65ex {\rm l}}}
\def\1{{1\kern-.65ex {\rm l}}}
\def\ap{{\alpha'}}

\def\Im{\mathop{\mathrm{Im}}\nolimits}

\def\ket#1{{|{#1}\rangle}}

\def\at{{\widetilde{a}}}
\def\bt{{\widetilde{b}}}

\def\Kt{{\widetilde{K}}}

\def\phit{{\widetilde{\phi}}}
\def\psit{{\widetilde{\psi}}}

\def\cA{{\cal A}}

\def\cC{{\cal C}}

\def\cF{{\cal F}}

\def\cH{{\cal H}}
\def\cI{{\cal I}}

\def\cM{{\cal M}}
\def\cN{{\cal N}}
\def\cO{{\cal O}}

\def\cQ{{\cal Q}}

\def\bbR{{\mathbb{R}}}

\def\bbZ{{\mathbb{Z}}}

\newcount\hour \newcount\minute
\hour=\time \divide \hour by 60
\minute=\time
\def\now{%
\ifnum \hour<13
  \ifnum \hour=0 \advance \hour by 12 \number\hour:\else \number\hour:\fi%
     \ifnum \minute<10 0\fi%
     \number\minute%
\ A.M.%
\else \advance \hour by -12 \number\hour:%
  \ifnum \minute<10 0\fi%
  \number\minute%
  \ P.M.%
\fi%
}

\makeatother

\begin{document}

\baselineskip=18pt  
\numberwithin{equation}{section}  

\renewcommand{\headrulewidth}{0pt}
%


%
%


\thispagestyle{empty}

\vspace*{-2cm} 
\begin{flushright}
YITP-21-47\\
\end{flushright}

\vspace*{2.5cm} 
\begin{center}
 {\LARGE Interpolating between multi-center microstate geometries}\\
 \vspace*{1.7cm}
 Masaki Shigemori\\
 \vspace*{1.0cm} 
Department of Physics, Nagoya University\\
Furo-cho, Chikusa-ku, Nagoya 464-8602, Japan\\
and\\
Center for Gravitational Physics,\\
Yukawa Institute for Theoretical Physics, Kyoto University\\
Kitashirakawa Oiwakecho, Sakyo-ku, Kyoto 606-8502, Japan
\end{center}
\vspace*{1.5cm}

\noindent
We study interpolation between two multi-center microstate
geometries in 4d/5d that represent Lunin-Mathur geometries with circular
profiles.  The interpolating solution is a Lunin-Mathur geometry with a
helical profile, and is represented by a 2-center solution with a
codimension-2 source.  The interpolating 2-center solution exhibits
interesting features such as some of the charges being delocalized, and
some of the charges getting transferred from the codimension-2 center to
the other, codimension-3 center as the interpolation proceeds.
We also discuss the spectral flow of this entire process and speculate
on the relevance of such solutions to understanding general microstates
of 3-charge black holes.

\newpage
\setcounter{page}{1} 





\section{Introduction and summary}

String theory contains various extended objects---branes---and allows
configurations with multiple such branes bound together by their
gravitational and gauge interactions. In supergravity, such
configurations are realized by solutions with multiple centers
representing branes. Among such multi-center solutions in supergravity,
an important class of solutions is the one in four and five dimensional
supergravity that is supersymmetric and characterized by harmonic
functions in $\bbR^3$ \cite{Behrndt:1997ny,Gauntlett:2002nw,Bates:2003vx,Bena:2004de,Gauntlett:2004qy,Bena:2005va,Meessen:2006tu}. 
They have various applications such as the attractor mechanism
\cite{Ferrara:1995ih,Strominger:1996kf,Ferrara:1996dd,Ferrara:1996um,Moore:2004fg,Kraus:2005gh,Larsen:2006xm},
split attractor flows and wall crossing
\cite{Denef:2000ar,Denef:2001ix,Moore2010pitp,Denef:2007vg,Bates:2003vx},
and microstate geometries
\cite{Bena:2005va,Berglund:2005vb,Heidmann:2018vky}.

In this note we will call these solutions ``harmonic solutions'',
because their construction heavily relies on harmonic functions.  In
most literature those harmonic solutions are assumed to have sources of
codimension three, but they can also have codimension-2
sources.\footnote{Codimension-1 singularities are also possible,
although we do not consider them in this note.}  Codimension-2 sources
can be produced by the supertube transition \cite{Mateos:2001qs} and
harmonic functions will have non-trivial monodromies around a curve
in~$\bbR^3$.  Some examples of codimension-2 harmonic solutions were
studied in \cite{Park:2015gka,Fernandez-Melgarejo:2017dme}.

In this note, we extend the examples of codimension-2 harmonic
solutions by studying the Lunin-Mathur geometries \cite{Lunin:2001jy,
Lunin:2002iz} in the framework.  The Lunin-Mathur geometries are smooth,
horizonless solutions of type IIB supergravity in $\bbR_t\times
\bbR^4\times S^1_y\times T^4$ and represent microstates of the D1-D5
2-charge system.  Ignoring $T^4$ directions they can be regarded as
solutions of 6d supergravity.  The geometries are parametrized by 
profile functions which
describe the shape in $\bbR^4$ of the worldvolume of a Kaluza-Klein
monopole (KKM)  produced by the supertube transition of D1-
and D5-branes.  For some special choices of the profile functions, via
duality transformations, the Lunin-Mathur geometries can be described by
harmonic solutions with codimension-3 sources.  Here we consider more
general profile functions, for which the harmonic functions have
codimension-2 sources as well.\footnote{The duality transformations at the
supergravity level involve smearing, and therefore the profile function
in $\bbR^4$ gets smeared along a direction $\psi$ along which T-duality
is taken.  This is unlike the formulation of \cite{Niehoff:2013kia} which
generalizes the harmonic function to depend on $\psi$.  }

Our purpose here is to develop techniques to construct harmonic
solutions with co\-di\-men\-sion-2 sources.  The harmonic solution
involves harmonic functions commonly denoted by $(V,K^I,L_I,M)=:\cH$ where $I=1,2,3.$  The
construction proceeds in layers, in that one constructs the harmonic
functions in the order of $V,K^I,L_I,M$.  In the previous work
\cite{Park:2015gka, Fernandez-Melgarejo:2017dme}, solutions with
trivial~$V$ (i.e.~$V=1$) were studied.  In the current note, we will
extend this to solutions with non-trivial~$V$.  This is a step forward
for constructing the most general codimension-2 harmonic solutions.

\bigskip
In the remainder of this section, we introduce the setup and summarize
our main findings.

The Lunin-Mathur geometry is parametrized by profile functions
$g_m(\lambda)$, $m=1,2,3,4$ in~$\bbR^4$, where $g_m(\lambda+L)=g_m(\lambda)$ with
$L$ a constant.  One of the simplest Lunin-Mathur geometry is given by
the following profile:
\begin{align}
 g_1(\lambda)+i g_2(\lambda)=ae^{i k\Omega\lambda},\qquad
 g_3(\lambda)+i g_4(\lambda)=0,
\label{profile_circular0}
\end{align}
where $a>0$, $k\in\bbZ_{>0}$ and $\Omega={2\pi/L}$. This is a circle in
the $x_1$-$x_2$ plane (and a point at the origin in the $x^3$-$x^4$ plane).  Using the Hopf fibration, we can project this
$\bbR^4$ profile onto $\bbR^3$ in which the harmonic functions $(V,K^I,L_I,M)$ of the
harmonic solution live. If we take the Hopf fiber direction $\psi$ to be
the same as the circle direction of the profile \eqref{profile_circular0},
the harmonic functions  are given by
\begin{align}
\begin{gathered}
 V={1\over r}, \qquad K^1=K^2=0,\qquad 
 K^3={Q_5\Omega k\over 2}\left({1\over \Sigma}-{1\over r}\right),\\
 L_1={Q_1\over 4\Sigma},\qquad
 L_2={Q_5\over 4\Sigma},\qquad
 L_3=1,\qquad M={Q_5\Omega k \at\over 4\Sigma}.
\label{VKLM_supertube}
\end{gathered}
\end{align}
Here the coordinates of $\bbR^3$ are $\yv=(y_i)$, $i=1,2,3$, and we
defined $r\equiv |\yv|$, $\Sigma\equiv|\yv-{\bf \at}|$, ${\bf \at}\equiv
(0,0,-\at)$, $\at={a^2/4}$.  These harmonic functions
\eqref{VKLM_supertube} have codimension-3 singularities at the origin
$\yv=0$ ($r=0$) and at a point on the negative $y_3$ axis, $\yv={\bf
\at}$ ($\Sigma=0$).   The singularities in $L_1,L_2$ at
$\Sigma=0$ correspond to the D1 and D5-brane charges that we start with,
while the singularity in $K^3$ at $\Sigma=0$ corresponds to the KKM
charge of the D1-D5 supertube along the Hopf fiber direction $\psi$.  To
avoid possible confusion, in the following, we refer to this charge
appearing as a codimension-3 source in $K^3$ as the \emph{D4 charge},
borrowing its dual type IIA interpretation.  This solution has a $U(1)$
rotational symmetry about the $y_3$ axis.

The need for the $1/r$ term in $K^3$ is not so obvious from the
viewpoint of the harmonic function, but this is what one gets from
reducing the Lunin-Mathur solution.  As we will see in the main text,
this is necessary for the gauge field in the Lunin-Mathur solution,
which involves $V^{-1}K^3$, to vanish at infinity.  Alternatively, we
can regard this $1/r$ term as coming from the ``gauge transformation''
\cite{Bena:2005ni} of harmonic solutions, under which physical fields
are invariant:
\begin{align}
\begin{split}
  V&\to V,\qquad K^I\to K^I+c^I V,
\end{split}
\label{gauge_transf0}
\end{align}
with $c^I$ arbitrary constants (here we are only showing the $V,K^I$
part; for the full expression including $L_I$ and $M$, see
\eqref{gauge_transf}).  It is clear that by choosing $c^3$ appropriately
we can change the coefficient of the $1/r$ term in $K^3$ as we want.

\begin{figure}[htb]
\begin{center}
\begin{tabular}{cccc}
  \begin{tikzpicture}[scale=0.65]
  \draw[-latex] (-2.5,0) -- (2.5,0) node [below] {$y_1$};
  \draw[-latex] (0,-4.5) -- (0,3.5) node [above]{$y_3$};

  \draw[dashed] (0,-0.7) circle (1.8 and 2);

  \fill[black!50!green] (0,0) circle (0.1) node [above left] {$-k$}; 

  \fill[magenta] (0,-2.7) circle (0.1) node [below right] {$k$};
 \end{tikzpicture}
 &
  \begin{tikzpicture}[scale=0.65]
  \draw[-latex] (-2.5,0) -- (2.5,0) node [below] {$y_1$};
  \draw[-latex] (0,-4.5) -- (0,3.5) node [above]{$y_3$};

  \draw[dashed] (0,-0.7) circle (1.8 and 2);

  \fill[black!50!green] (0,0) circle (0.1) node [above left] {$-k$};

  \draw[red,thick,decorate,decoration={zigzag,amplitude=1,segment length=4}]
     (-1,-2.35) -- +(2,0);
  \draw[red,latex-] (0.3,-2.35) -- +(.25,-.5) node 
     [below,xshift=15,yshift=3] {$\tfrac{(k+k')b^2}{a^2+b^2}$};

  \fill[blue] (-1,-2.35) circle (0.1);
  \fill[blue] (1,-2.35) circle (0.1);
  \draw[blue,latex-] (-1,-2.45) -- +(-0.25,-0.5) node
     [below,xshift=-3,yshift=3] {$\tfrac{ka^2-k'b^2}{a^2+b^2}$};
 \end{tikzpicture}
 &
  \begin{tikzpicture}[scale=0.65]
  \draw[-latex] (-2.5,0) -- (2.5,0) node [below] {$y_1$};
  \draw[-latex] (0,-4.5) -- (0,3.5) node [above]{$y_3$};

  \draw[dashed] (0,-0.7) circle (1.8 and 2);
  \fill[black!50!green] (0,0) circle (0.1) node [below left] {$k'$};

  \draw[red,thick,decorate,decoration={zigzag,amplitude=1,segment length=4}] (-1,0.95) -- +(2,0);
  \draw[red,latex-] (0.3,1.05) -- +(.25,.5) node 
     [above,xshift=13,yshift=-3] {$-\tfrac{(k+k')a^2}{a^2+b^2}$};

  \fill[blue] (-1,.95) circle (0.1);
  \fill[blue] ( 1,.95) circle (0.1);
  \draw[blue,latex-] (-1,1.05) -- +(-0.25,0.5) node
     [above,xshift=-3,yshift=-3] {$\tfrac{ka^2-k'b^2}{a^2+b^2}$};

 \end{tikzpicture}
&
  \begin{tikzpicture}[scale=0.65]
  \draw[-latex] (-2.5,0) -- (2.5,0) node [below] {$y_1$};
  \draw[-latex] (0,-4.5) -- (0,3.5) node [above]{$y_3$};

  \draw[dashed] (0,-0.7) circle (1.8 and 2);

  \fill[black!50!green] (0,0) circle (0.1) node [below left] {$k'$};

  \fill[magenta] (0,1.3) circle (0.1) node [above right] {$-k'$};
 \end{tikzpicture}
 \\[2ex]
 (a) south pole limit &
 (b) ``southern'' case &
     (c) ``northern'' case &
     (d) north pole limit
\end{tabular}
\begin{quote}
  \caption{\sl The $\bbR^3$ profile and D4 charges of the ``helical''
solution, as we change the parameters $a,b$.  
(a): the south pole limit ($a>0,b=0$).  The profile is a point at the
south pole of a spheroid (dashed ellipse).  The purple dot represents the
D4 charge at $\Sigma=0$, while the green dot the $r=0$ center. The D4 
charge is shown next to each center in units of $Q_5\Omega /2$. 
(b) as we make $b$ nonzero, the profile goes off the south pole and
becomes a curve $\cC$ along a latitude line of the spheroid.  The blue
dots represent the location of $\cC$.  The magenta zigzag line between
them represents the branch cut coming from the multi-valuedness of $K^3$.
Delocalized D4 charges are distributed on it.
(c) As we increase $b$ and the profile $\cC$ goes up, the branch cut crosses the $r=0$ center.
Some of the delocalized D4 charge has been transferred to the $r=0$
center.
(d) In the north pole limit $(a=0)$, the profile collapses to a point at
the north pole.  See the main text for more detail.
\label{fig:spheroid0}}
\end{quote} 
\end{center}
\end{figure}
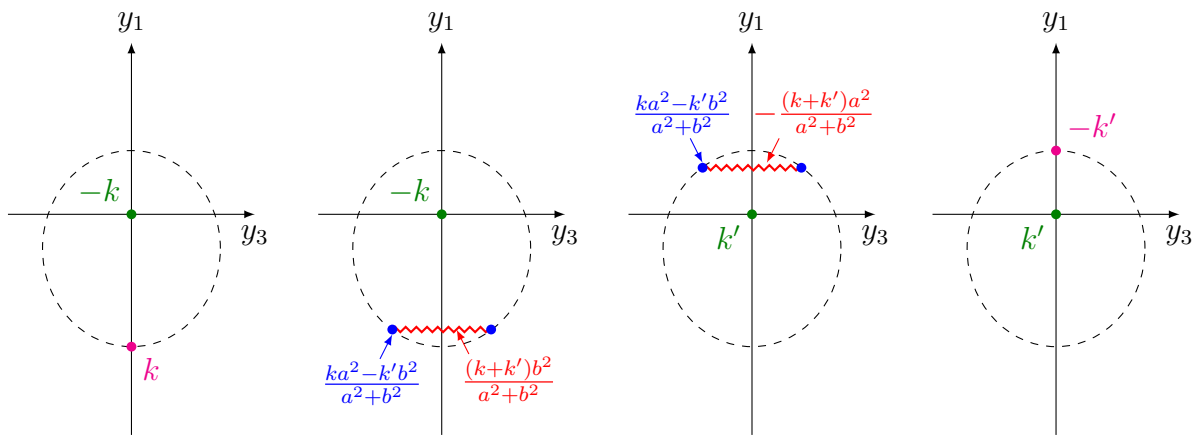

A more general profile---the object of main interest in the current
note---is
\begin{align}
 g_1(\lambda)+i g_2(\lambda)=ae^{i k \Omega\lambda},\qquad
 g_3(\lambda)+i g_4(\lambda)=be^{-i k'\Omega\lambda},
\label{profile_helix0}
\end{align}
where $a,b\ge 0$ and $k,k'\in\bbZ_{>0}$.  This is like a helix, going in
circles in both $x_1$-$x_2$ and $x_3$-$x_4$ directions, with pitches
$k,k'$.  When $b=0$, this reduces to \eqref{profile_circular0}.  For $b>
0$, the $\psi$ direction is not an isometry of the Lunin-Mathur solution, but we can still reduce it to $\bbR^3$ after smearing
it along the $\psi$ direction.  The projected profile in $\bbR^3$ is
like a latitude line on the globe.  
The explicit harmonic functions can
be found in section \ref{sec:codim-2_LM}. As we will see in the main text, the parameters
$a,b$ are not arbitrary but constrained to satisfy
\begin{align}
 a^2k^2+b^2k'^2=\text{const}={Q_1\over  Q_5\Omega^2}.\label{ghfp19Apr21}
\end{align}
Let's say we start with $a>0,b=0$, which corresponds to
\eqref{profile_circular0}, and increase $b$
satisfying~\eqref{ghfp19Apr21}, finally ending with $a=0,b>0$.  The
projected profile in the $\yv$ space starts as a point at the ``south
pole'' of a spheroid (at $a>0,b=0$) and, as we increase $a$, becomes a
latitude line.  As we go up the spheroid from the south toward the
north, the latitude line gets larger and then smaller, finally
collapsing to a point at the ``north pole'' (at $a=0,b>0$); see the
purple and blue dots in Figure~\ref{fig:spheroid0}.
When $a,b>0$, the latitude line $\cC$ is a codimension-2 singularity,
around which $K^3$ has an additive a monodromy.  This singularity
represents the KKM charge of the D1-D5 supertube along $\cC$.  Let us
refer to this charge appearing as a codimension-2 source in $K^3$ as the
\emph{NS5 charge}, borrowing its dual IIA interpretation.  This is the
most general Lunin-Mathur solution reduced to 4d that preserves the $U(1)$ rotational symmetry
about the $y_3$ axis.

Even for $a,b>0$ we have the D4 charge (KKM charge along $\psi$),
appearing as codimension-3 sources in $K^3$.  It is interesting to see
what happens to the D4 charge as we go from the south pole to the north
pole.  Being a dipole charge, the D4 charge is not conserved from the 5d
viewpoint, because the Hopf fiber that it is wrapping is contractible in
$\bbR^4$.  However, from the 4d viewpoint, the D4 charge is an ordinary
(monopole) charge as any other charges and is conserved.  This is
possible because the origin of $\bbR^4$, which is a D6-brane at $r=0$
in the IIA interpretation, can also carry D4 charges;
as the profile moves on the spheroid, some of the D4 charge can be
transferred to the D6 center at $r=0$, so that the total D4 charge is
conserved.  This is the same mechanism as in
\cite{Gregory:1997te}, where they studied how the charge of a
fundamental string wrapping the special circle of a KKM background can
be conserved, even though the circle is contractible.

Here we stop to note that there are multiple notions of charges
\cite{Marolf:2000cb} and the one we are talking about is the Page
charge, which is conserved, localized, and quantized, but changes under
large gauge transformations.  For harmonic solutions, the D4 Page charge
is measured by \cite{Park:2015gka}
\begin{align}
-{1\over 4\pi} \int_{M} *_3 dK^3,
\end{align}
where $*_3$ is the Hodge star for flat $\bbR^3$ and $M$ is a Gaussian
surface.  If $K^3$ is single-valued, this simply picks up the coefficients of
poles in $K^3$.

In the present setting, charge conservation works as follows.  Because
of the non-single-valuedness of $K^3$, in addition to the charge source
appearing as a pole in the harmonic function, there also are
``delocalized'' charges that can be measured by the Gaussian surface
surrounding the disk whose boundary is $\cC$.\footnote{This delocalized
charge is an example of ``Cheshire charges'' that can appear in the
presence of monodromies \cite{Schwarz:1982ec}.}$^,$\footnote{ The
delocalized charge here corresponds to the fundamental string charge
carried by the KKM background in the context of \cite{Gregory:1997te}.}
This disk is the branch cut across which $K^3$ is discontinuous.  As
$\cC$ moves up on the spheroid, when the center at $r=0$ crosses the
branch cut, the delocalized charge gets transferred to the $r=0$ center.
This process is described in Figure~\ref{fig:spheroid0}.  For $a>0,b=0$
(south pole), there is a non-vanishing D4 charge but no NS5 charge
(Figure~\ref{fig:spheroid0}(a)).  As we move off the south pole, we
start to have a non-vanishing NS5 charge along a now finite ring $\cC$.
The D4 charge is divided into a localized part on curve $\cC$ and a
delocalized part distributed over the branch cut
(Figure~\ref{fig:spheroid0}(b)).  After the branch cut has passed the
$r=0$ center, the delocalized charge has changed by the amount
transferred to the $r=0$ center (Figure~\ref{fig:spheroid0}(c)).
Finally, when we reach the north pole, curve $\cC$ shrinks to a point
and we are left with a different amount of D4 charge
(Figure~\ref{fig:spheroid0}(d)).

So, the trivial-looking process of continuously changing the
Lunin-Mathur profile in 6d leads to a non-trivial process in the 4d harmonic
solution, in which the D4 charge gets transferred to the D6 center by
way of the dimension-2 NS5-brane.  The significance of this is further
discussed in section~\ref{sec:cft} in relation to spectral flow in the
dual CFT\@.  Generalizations of such processes of charge transfer are
expected to realize more general topology-changing processes that
presumably play an important role in understanding the microscopic
physics of black holes, as will be discussed in
section~\ref{sec:discussions}.

No solutions or geometries presented in this note are essentially new;
the purpose here is to look at them from a new viewpoint and
re-interpret them, potentially as a basis on which to construct new
solutions.

In the rest of the note, we will give more details of the picture
explained above, and also discuss related matters.  First, in
section~\ref{sec:setup}, we explain harmonic solutions, the Lunin-Mathur
geometries, and the relation between them.  Then, in
section~\ref{sec:codim-2_LM}, we derive the harmonic functions that
corresponds to the Lunin-Mathur geometry for the ``helical'' profile
\eqref{profile_helix0}, and examine in detail the process of D4-charges
getting transferred to the D6 center.  We will also discuss various
other matters such as how the constraint \eqref{ghfp19Apr21} is derived
from a no-CTC (closed timelike curve) condition.  In
section~\ref{sec:from_4+1}, we confirm the harmonic functions derived in
section~\ref{sec:codim-2_LM} from the 6d Lunin-Mathur geometry side.
In section~\ref{sec:cft}, we discuss the dual CFT
perspective of the whole process.  We will also consider the spectral flow and fractional spectral flow
of our solution, in both CFT and gravity.
Finally, in
section~\ref{sec:discussions}, we discuss the implication of the result
obtained in this note.
Some details of the computation in the main text can be found in the
Appendix.

\section{Setup}
\label{sec:setup}

\subsection{Harmonic solutions}

\label{ss:harm_sol}

Here we give a brief review of the harmonic solution, which represents
multi-center black-hole/ring solutions in 4d/5d.  Our purpose here
is to introduce notation; for further detail, see
\cite{Behrndt:1997ny,Gauntlett:2002nw,Bates:2003vx,Bena:2004de,Gauntlett:2004qy,Bena:2005va,Meessen:2006tu}
 (for
solutions with codimension-2 sources see also
\cite{Park:2015gka,Fernandez-Melgarejo:2017dme}).

The most general supersymmetric solutions of ungauged 5d $\cN=1$
supergravity with vector multiplets have been classified in
\cite{Gutowski:2004yv} (see also \cite{Gauntlett:2002nw, Bena:2004de,
Gutowski:2005id}). When one applies this result to M-theory compactified
on $T^6=T^2_{45}\times T^2_{67}\times T^2_{89}$ (the so-called STU
model) and further assumes a tri-holomorphic $U(1)$ symmetry
\cite{Gauntlett:2004qy}, the general supersymmetric solution corresponds
to the following 11d fields:
\begin{align} 
\label{Msol}
\begin{aligned}
ds_{11}^2
&=
-Z^{-2/3}(dt+k)^2
+Z^{1/3}ds_{\mathrm{GH}}^2
+Z^{1/3}\left(Z_1^{-1}dx_{45}^2
+Z_2^{-1}dx_{67}^2+Z_3^{-1}dx_{89}^2\right) 
\, ,
\\[.5ex]
\cA_3
&=
\left(B^I - Z_I^{-1}(dt+k)\right)\wedge  J_I
\, ,
\quad
J_1 \equiv dx^4\wedge dx^5\,,~
J_2 \equiv dx^6\wedge dx^7\,,~
J_3 \equiv dx^8\wedge dx^9\,,
\end{aligned}
\end{align}
where $I=1,2,3$; $Z\equiv Z_1 Z_2 Z_3$; and $dx_{45}^2\equiv
(dx^4)^2+(dx^5)^2$ etc.
Supersymmetry implies that all fields in \eqref{Msol} are written in
terms of 3D harmonic functions
\begin{align}
 \cH\equiv (V,K^I,L_I,M)
\end{align}
as follows.  First,
 $ds_{\mathrm{GH}}^2$ is a 4-dimensional metric of a
Gibbons-Hawking space given by
\begin{align}
ds_{\mathrm{GH}}^2
=
V^{-1}(d\psi+A)^2
+V d\yv^2
\, , \qquad 
 \psi\cong \psi+4\pi\,,\qquad
\yv=(y_1,y_2,y_3)
\, .
\label{GHmetric}
\end{align}
The 1-form $A$ and the scalar $V$ depend on
the coordinates $\yv$ of the $\mathbb{R}^3$ base and satisfy
\begin{align}
dA= *_3\, dV
\, ,
\end{align}
where $*_3$ is the Hodge dual in flat $\mathbb{R}^3$.  We generally
denote 3d vectors by boldface letters, such as $\yv=(y_i)$,
$i=1,2,3$. The rest of the fields are:
\begin{subequations} 
 \begin{align}
 B^I &= V^{-1} K^I (d\psi+ A)+ \xi^I
 \, ,
 \qquad\qquad
 d \xi^I
 =
 - *_3 dK^I
 \,   ,\label{BI,xiI}\\
 Z_I 
 &=
 L_I+\frac12 C_{IJK} V^{-1} K^J K^K
 \, ,
 \\
 k 
 &=
 \mu (d\psi + A) 
 +\omega
 \, ,
 \label{k1form}
 \\
 \mu 
 &=M+
 \frac12 V^{-1} K^I L_I
 +\frac16 C_{IJK} V^{-2} K^I K^J K^K
 \, ,
\label{hljz8Apr21}
 \end{align}
\end{subequations}
where $C_{IJK}=|\epsilon_{IJK}|$.  All fields are assumed to depend only
on $\yv$ and not $\psi$ or $T^6$ coordinates.  The physical fields in
the solution, such as $Z_I,\mu$, are invariant under the ``gauge transformation''
\cite{Bena:2005ni}
\begin{align}
\begin{split}
  V&\to V,\qquad K^I\to K^I+c^I V,\\
 L_I&\to L_I-C_{IJK}c^J K^K-{1\over 2}C_{IJK}c^J c^K V,\\
 M&\to M-{1\over 2}c^I L_I+{1\over 12}C_{IJK}(c^I c^J c^K V+3c^I c^J K^K)
\end{split}
\label{gauge_transf}
\end{align}
where $c^I$ are arbitrary constants.  In the 5d setup that includes
$\psi$ as a coordinate, this is a gauge transformation for which the
gauge transformation parameter depends on $\psi$ (it shifts the
coefficient of $d\psi$ in $B^I$).  However, in the 4d context, such $\psi$
dependence is not allowed and the transformation \eqref{gauge_transf}
does change the charge \cite{Hanaki:2007mb}.

As already stated, supersymmetry
requires that $\cH=(V,K^I,L_I,M)$ be harmonic functions in $\bbR^3$:
\begin{align}
 \Lap V= \Lap K^I= \Lap L_I= \Lap M=0,\qquad
 \Lap \equiv \partial_i \partial_i.  
\end{align}
The 1-form $\omega$ satisfies
\begin{align}
 *_3 d \omega 
 =
 V d M - M dV +\frac12(	K^I dL_I	- L_I d K^I	)
 \, .
 \label{sdomega}
\end{align}
Applying $d\,*_3$ on this equation we obtain the ``integrability condition''
\cite{Denef:2000nb} or the ``bubble equation'' \cite{Bena:2005va}
\begin{align}
0 =
 V {\Lap M}- M {\Lap V}
 +\frac12 (K^I \Lap L_I	- L_I \Lap K^I	)
\, .
\label{integrability}
\end{align}
Although the functions $V,K^I, L_I, M$ are harmonic, they have sources
and the right-hand side of \eqref{integrability} has delta-function
singularities.  The integrability condition requires all such
singularities cancel \cite{Bena:2005va}.\footnote{For a discussion on the
physical meaning of this condition for codimension-2 sources see \cite{Fernandez-Melgarejo:2017dme}.}

\bigskip
Reducing the 11D solution \eqref{Msol} on the $\psi$ circle $S^1_\psi$, we obtain the
following supersymmetric solution of type IIA supergravity:
\begin{align}
\begin{split}
 ds_{10,\mathrm{str}}^2
 &=
 -\frac{1}{\sqrt{\cQ\,}}(dt+\omega)^2
 +\sqrt{\cQ\,} \, d\xv^2
 + \frac{\sqrt{\cQ\,}}{V}\left(Z_1^{-1}dx_{45}^2
 +Z_2^{-1}dx_{67}^2+Z_3^{-1}dx_{89}^2\right)
 \, ,
 \\
 e^{2\Phi}
 &=
 \frac{\cQ^{3/2}}{V^{3} Z}
 \, ,\qquad
 B_2=\left({K^I\over V}-{\mu\over Z_I }\right) J_I\, ,
\end{split}
\label{IIAfield}
\end{align}
where $ds^2_{\rm 10,str}$ is the string-frame metric and
$\cQ \equiv V(Z-\mu^2 V)$.
There are also RR potentials whose explicit form can be found
e.g.~in \cite[App.~E]{Park:2015gka} and \cite{DallAgata:2010srl}.
The complexified K\"ahler modulus associated with $T_{89}^2$ is defined by
\begin{align}
 \tau^3={R_8R_9\over \ap}\left(B_{89}+i\sqrt{\det G_{ab}}\right)=
 {R_8R_9\over \ap}\left[\left({K^3\over V}-{\mu\over Z_3 }\right)
 +i{\sqrt{\cQ}\over Z_1 V}\right] ,
\label{def_tau^I}
\end{align}
where 
$a,b=8,9$ and
$R_i$ are the radii of the $x^i$ directions, $i=4,\dots,9$.
Under the gauge transformation~\eqref{gauge_transf},
this transforms as 
\begin{align}
 \tau^3\to  \tau^3 + {R_8 R_9\over \ap}c^3.
\end{align}
The moduli $\tau^1$ and $\tau^2$ for $T^2_{45}$ and $T^2_{67}$ are
defined similarly, and the transformation under the gauge
transformation~\eqref{gauge_transf} is similar.

\subsubsection*{Codimension-3 sources}

The harmonic functions $\cH=(V,K^I,L_I,M)$ can have sources that represent
branes, which are D-branes in the IIA setup.  
If one assumes that all sources are of
codimension 3, the harmonic functions can be written as
\begin{align}
\cH=h+\sum_{p=1}^N {\Gamma_p\over |\yv-\av_p|},
\label{codim3harm}
\end{align}
where $\av_p\in \mathbb{R}^3$ ($p=1,\dots,N$) specify the location of
the codimension-3 sources where the harmonic functions become singular,
and $h,\Gamma_p$ are constants.

The codimension-3 sources in the harmonic functions \eqref{codim3harm}
represent branes in string/M-theory.
For example, in the type IIA picture
\eqref{IIAfield}, 
the dictionary  between the singularities in the
harmonic functions and the D-brane sources is
\cite{Bates:2003vx}
\begin{align}
V
\leftrightarrow
\text{D6(456789)}
\, ,
\quad
\begin{array}{l}
K^1\leftrightarrow \text{D4(6789)}
\\[5pt]
K^2\leftrightarrow \text{D4(4589)}
\\[5pt]
K^3\leftrightarrow \text{D4(4567)}
\end{array}
\, ,
\quad
\begin{array}{l}
L_1\leftrightarrow \text{D2(45)}
\\[5pt]
L_2\leftrightarrow \text{D2(67)}
\\[5pt]
L_3\leftrightarrow \text{D2(89)}
\end{array}
\, ,
\quad
M \leftrightarrow \text{D0}
\, .
\label{singIIAbrn}
\end{align}
The D-branes are partially wrapped on $T^6$ as indicated here and appear in
4D as pointlike (codimension-3) objects sourcing the harmonic functions.
When lifted to M-theory, D4 becomes M5 wrapping $S^1_\psi$ and D2
becomes M2.  D0 becomes momentum (P) along $S^1_\psi$ while D6 becomes
Kaluza-Klein monopole (KKM) with $S^1_\psi$ being its special circle.

Depending on how to choose the base space \eqref{GHmetric}, the harmonic
solution can describe 4d solutions or 5d solutions.  For example, 
take
\begin{align}
 V={1\over r}, \qquad K^I=0,\qquad
 L_I=1+{Q_I\over 4r},\qquad M=0,\qquad r\equiv |\yv|.\label{ieyh21May21}
\end{align}
In type IIA, this describes a 4d black hole made of three stacks of
D2-branes and a D6-brane.
Meanwhile, we can interpret this also as a 5d black hole because, for
this $V$, the base space~\eqref{GHmetric} describes flat $\bbR^4$ via a
Hopf fibration.  Namely, the metric for flat $\bbR^4$ with
coordinates~$x_m$, $m=1,2,3,4$ can be written as
\begin{align}
\begin{split}
  ds^2_{\bbR^4}
 &=dx_m dx_m
 =V^{-1}(d\psi+A)^2+V\bigl(dr^2+r^2(d\theta^2+\sin^2\!\theta\,d\phi^2)\bigr),
 \\
 V&={1\over r},\qquad A=(1+\cos\theta)d\phi,
\end{split}
 \label{Hopf_fib}
\end{align}
where
\begin{align}
 x_1+ix_2&=2\sqrt{r}\,\sin{\theta\over 2}\,e^{i{\psi\over 2}},\qquad
 x_3+ix_4=2\sqrt{r}\,\cos{\theta\over 2}\,e^{i({\psi\over 2}+\phi)}.
\label{xm_Hopf_fib}
\end{align}
with $\psi\cong \psi+4\pi$, $\phi\cong\phi +2\pi$.  The Cartesian
coordinate $\yv$ in the 3d base is
\begin{align}
 y_1+iy_2=r\sin\theta\, e^{i\phi},\qquad
 y_3=r\cos\theta.\label{def_y}
\end{align}
Lifted to M-theory along the $\psi$ direction, this harmonic solution
becomes a 5d black hole made of three stacks of M2-branes, the D6-brane
becoming the origin of the $\bbR^4$.  This 5d black hole can be dualized
(using the duality of Appendix \ref{app:duality}) into the original Strominger-Vafa black
hole \cite{Strominger:1996sh} in a type IIB frame, where the M2/D2
charges are mapped into D1, D5, and P charges.

\subsubsection*{Codimension-2 sources}

Harmonic functions can have other kinds of source.  They can have a
singularity along a curve in $\bbR^3$ and have non-trivial monodromy
around it.  We refer to such singularities as codimension-2
sources. Note that this is genuinely
different from the codimension-3 source discussed above; we can have a
continuous distribution of codimension-3 sources along a curve, but we
will still refer to them as codimension-3 sources.

One situation for codimension-2 sources to appear is when branes undergo
a supertube transition \cite{Mateos:2001qs}, gaining dimension (or losing codimension). For
example, the following supertube transition is possible:
\begin{align}
 \text{D2}(45) + \text{D2}(67)
 \xrightarrow{\text{puff out}} 
 \text{NS5}(4567\lambda)+\text{P}(\lambda)
 ,\label{D2+D2->NS5}
\end{align}
where in the final configuration there is an NS5-brane along internal
(4567) directions and a closed curve $\cC$ in $\bbR^3$ parametrized by
$\lambda$, and momentum along $\lambda$.  By looking at the expression for $B_2$ in \eqref{IIAfield}, we
see that, if $V^{-1}K^3$ has a monodromy around $\cC$, there will be an
NS5-brane along $\cC$ (and 4567).  (More precisely it is
$V^{-1}K^3-Z_3^{-1}\mu$ that must be monodromic, but in the current
note we assume that $Z_I,\mu$  appearing in the metric are single-valued
and therefore it is the monodromy of $V^{-1}K^3$ that is relevant.)  As
it turns out, in this case $M$ also becomes monodromic.  In general,
given codimension-2 brane sources, which harmonic functions to become
monodromic is a non-trivial matter that depends on the physical
situation in question.

\subsection{Lunin-Mathur geometries}
\label{ss:LM_geom}

The Lunin-Mathur geometry \cite{Lunin:2001jy, Lunin:2002iz} is a
solution of type IIB supergravity in $\bbR_t\times \bbR^4_{1234}\times
S^1_y\times T^4_{6789}$ and represents microstates \cite{Rychkov:2005ji,
Krishnan:2015vha} of the D1-D5 2-charge system.  The solution is
parametrized by profile functions $g_m(\lambda)$, $m=1,2,3,4$ satisfying
$g_m(\lambda+L)=g_m(\lambda)$, which describe the shape inside $\bbR^4$
of the KKM dipole produced by the supertube transition
of D1$(y)$ and D5($y$6789) branes.\footnote{We do not consider $g_A$
with $A\ge 5$ \cite{Kanitscheider:2007wq}, which describe other possible
dipole charges produced by the supertube transition.}  Just as
\eqref{D2+D2->NS5},  we can describe this by the following diagram:
\begin{align}
 \text{D1}(y) + \text{D5}(y6789)
  \xrightarrow{\text{puff out}} 
 \text{KKM}(6789\lambda,y)+\text{P}(\lambda),
\label{D1+D5->KKM+P}
\end{align}
where $\text{KKM}(6789\lambda,y)$ denotes the KKM dipole with $y$ being
the special circle.

The explicit form
of the 10d string-frame metric of the Lunin-Mathur geometry is
\begin{subequations}\label{ansatzSummary}
 \begin{align}
d s^2_{10} & = -{2\over \sqrt{Z_1 Z_2}}(dv+\betaLM )
  \Bigl(du+\omegaLM+{\cF\over 2}(dv+\betaLM)\Bigr)
  +\sqrt{Z_1Z_2}\,ds^2_{\bbR^4}+ \sqrt{\frac{Z_1}{Z_2}}\,ds^2_{T^4},\label{LM_geom_metric}
\end{align}
\end{subequations}
where
\begin{subequations}
\label{LM_geom_funcs}
\begin{gather}
 Z_1 =  \frac{Q_5}{L} \int_0^{L} d\lambda \frac{|\partial_\lambda \vec g(\lambda )|^2}{|\vec x -\vec g(\lambda )|^2},\qquad
 Z_2 =  \frac{Q_5}{L} \int_0^{L} \frac{d\lambda }{|\vec x -\vec g(\lambda )|^2},
 \\
 \ALM
 = - \frac{Q_5}{L} dx^j \int_0^{L} d\lambda  \frac{\partial_\lambda g_j(\lambda )}{|\vec x -\vec g(\lambda )|^2} ,\qquad  d\BLM = - *_4 d\ALM,
 \label{def_ALM}
 \\
 Q_1 =  \frac{Q_5}{L} \int_0^{L} d\lambda \,|\partial_\lambda \vec g(\lambda )|^2,\label{def_Q1}
 \\
 \betaLM = \frac{-\ALM+\BLM}{\sqrt{2}},\qquad\omegaLM = \frac{-\ALM-\BLM}{\sqrt{2}}\label{hegv8Apr21}
\end{gather}
\end{subequations}
and $ds^2_{\bbR^4}=dx_m dx_m$, $m=1,2,3,4$ is the flat $\bbR^4$ metric with
coordinates $\vec x=(x_m)$.
For later convenience, we have written the metric \eqref{LM_geom_metric}
in the general form given in \cite[Appendix E]{Giusto:2013bda} which
represents the 3-charge solution. In the 2-charge case we are
considering, we must set $\cF=0$.  Also, we have taken the decoupling
limit and dropped ``1'' from $Z_{1,2}$.  The coordinates $u,v$ are
related to time $t$ and the coordinate $y$ of the compact circle $S^1_y$
of radius $R_y$ as
\begin{align}\label{sptvw}
u = \frac{1}{\sqrt{2}}(t-y), \qquad v = \frac{1}{\sqrt{2}}(t+y).
\end{align}
The periodicity $L$ is related to $R_y$ as $L=2\pi Q_5/R_y$.  The
quantities $Q_1$, $Q_5$ are related to the quantized D1 and D5 numbers
$N_1$, $N_5$ by
\begin{align}
Q_1 = \frac{N_1 g_s \alpha'^3}{v_4}\,,\qquad Q_5 = N_5 g_s \alpha',
\label{Q1Q5_n1n5}
\end{align}
where $(2\pi)^4 v_4$ is the coordinate volume of $T^4$.  Finally, $*_4$
is the Hodge dual in flat $\bbR^4$ with coordinates $x_m$.

The simplest example of the profile, which has already been introduced
in the introduction, is a circle in the $x_1$-$x_2$ plane:
\begin{align}
 g_1+ig_2=ae^{i k \Omega \lambda},\qquad
 g_3+ig_4=0,
\label{profile_circular}
\end{align}
where $a>0$, $k\in\bbZ_{>0}$ and
\begin{align}
 \Omega\equiv{2\pi\over L}={R_y\over Q_5}.
\label{def_Omega}
\end{align}
Substituting this into \eqref{LM_geom_funcs}, we find
\begin{subequations}
\label{harm_func_circular}
\begin{gather}
 Z_1={Q_1\over 4\Sigma},\qquad Z_2={Q_5\over 4\Sigma},\label{Z1Z2_south_pole}\\
 Q_1=Q_5 a^2k^2\Omega ^2,\label{Q1_south_pole}\\
 \ALM=-{1\over 2}{Q_5 k \Omega}\left({s^2+a^2+w^2\over 4\Sigma}-1\right)d\phit,\\
 \BLM=+{1\over 2}{Q_5 k \Omega}\left({s^2-a^2+w^2\over 4\Sigma}-1\right)d\psit,
\end{gather}
\end{subequations}
where
\begin{align}
 \Sigma\equiv{1\over 4}\sqrt{[(s+a)^2+w^2][(s-a)^2+w^2]}
\label{def_Sigma_6d}
\end{align}
and we defined the polar coordinates $(s,\phit),(w,\psit)$ 
via
\begin{align}
 x_1+ix_2=se^{i\phit},\qquad
 x_3+ix_4=we^{i\psit},
\label{s_w_phit_psit}
\end{align}
with $s,w\ge 0$ and $\phit\cong \phit+2\pi$, $\psit\cong\psit+2\pi$.

\subsection{Relation to harmonic solutions}
\label{ss:rel_to_harm_sol}

By appropriate duality transformations (see Appendix \ref{app:duality}), we can
map the harmonic solutions in the M/type-IIA frame in
section~\ref{ss:harm_sol} into the type IIB D1-D5 frame in
section~\ref{ss:LM_geom}.  Here let us study the relation between the
fields in the Lunin-Mathur geometries and the harmonic functions in the
harmonic solutions.

As discussed in \cite{Bena:2008dw}, the harmonic solutions in the D1-D5 frame is
\begin{align}
 ds_{\rm IIB}^2&=-{1\over Z_3\sqrt{Z_1Z_2}}(dt+\kappa)^2+{Z_3\over \sqrt{Z_1Z_2}}(dz+A^{3})^2+
 \sqrt{Z_1Z_2}\,ds^2_{\rm GH} + \sqrt{Z_1\over Z_2}ds^2_{T^4},
\label{jdtd17Feb21}
 \\
 \kappa &=\mu (d\psi+A) +\omega,\qquad
 A^{3}=B^3-{1\over Z_3}(dt+\kappa),
\end{align}
where $B^3$ is the one given in \eqref{BI,xiI}.  On the other hand, the
first part (the first term) of the Lunin-Mathur solution
\eqref{LM_geom_metric} can be rewritten as
\begin{align}
 ds_{10}^2
 &=-{(dt-\ALM)^2\over Z_3\sqrt{Z_1 Z_2}}
 +{Z_3\over \sqrt{Z_1 Z_2}}\biggl(dy+dt+ \BLM-\ALM-{dt-\ALM\over Z_3}\biggr)^2
 +\cdots,\label{jdtp17Feb21}
\end{align}
where $Z_3\equiv 1-\cF/2$ (actually $Z_3=1$ in the D1-D5 case).  By
comparing \eqref{jdtd17Feb21} and \eqref{jdtp17Feb21}, we read off the
relation between the 1-forms $\ALM,\BLM$ of the Lunin-Mathur
geometry and the harmonic functions:
\begin{subequations} 
 \label{jtou17Feb21}
 \begin{align}
 -\ALM &~~\leftrightarrow~~ \kappa=\mu (d\psi +A)+\omega,\label{mfxm17Feb21}\\
 -\ALM+\BLM &~~\leftrightarrow~~ B^{3}=V^{-1}K^{3}(d\psi+A)+\xi^{3}\label{eqnb18Feb21}
 \end{align}
\end{subequations}
and $y+t \leftrightarrow z$.  With this dictionary, we can read off
harmonic functions that correspond to a Lunin-Mathur solution, and vice
versa.

Comparison of \eqref{LM_geom_metric} and \eqref{jdtd17Feb21} also means
that the flat $\bbR^4$ in which the D1-D5 profile $g_m(\lambda)$ lives
should be written as the Hopf fibration
\eqref{Hopf_fib}.\footnote{Another possibility is to compactify $\bbR^4$
trivially to $\bbR^3\times S^1_\psi$ by taking $V=1,A=0$.  Codimension-2
harmonic solutions obtained this way was discussed in
\cite{Park:2015gka}.}  Generically, the profile in $\bbR^4$ is going
along some curve $\cC$ in the $\bbR^3$ base, at the same time moving in
the $\psi$ fiber.  Because the harmonic solutions are independent of the
$\psi$ fiber, we must smear the profile along $\psi$.  The exception is
when the profile is purely in the $\psi$ direction and $\cC$ is a point
in the base.  One such exceptional example is the circular profile
\eqref{profile_circular}, for which $\cC$ corresponds to the point on
the $y_3$ axis, $r=a^2/4,\theta=\pi$ ($y_3=-a^2/4$).

We can see which harmonic functions get sources as follows.  The
$\lambda$ direction of the profile in \eqref{D1+D5->KKM+P} can now be
along $\psi$ or $\cC\subset\bbR^3$, so we have two kinds of
dipole charge that can be produced by the supertube transition:
\begin{align}
 \text{D1}(y) + \text{D5}(y6789)
 \xrightarrow{\text{puff out}} 
\begin{cases}
 \text{KKM}(6789\psi,y)+\text{P}(\psi)\\[.5ex]
 \text{KKM}(6789\cC,y)+\text{P}(\cC)
\end{cases}
\label{D1+D5->KKM2}
\end{align}
After the duality transformation of Appendix \ref{app:duality}, this is mapped
into the following process in~IIA:
\begin{align}
 \text{D2}(45) + \text{D2}(67)
 \xrightarrow{\text{puff out}} 
\begin{cases}
 \text{D4}(4567)+\text{D0}\\[.5ex]
 \text{NS5}(4567\cC)+\text{P}(\cC)  
\end{cases}
\label{D2+D2->D4+NS5}
\end{align}
The first line means that $K^3$ will have codimension-3 sources for
$\text{D4}(4567)$. The codimension-3 charge is continuously distributed
along $\cC$, and its local density is proportional to the ``speed'' of
the profile going along $\psi$.  In addition, $K^3$ will have
codimension-2 sources for $\text{NS5}(4567\cC)$; namely, $V^{-1}K^3$
will have a constant additive monodromy as we mover around $\cC$ and the
constant is proportional to the NS5-brane charge.  Also, $M$ will have
codimension-3 sources for $\text{D0}$, while momentum $\text{P}(\cC)$
will be encoded in the 1-form $\omega$.

We can confirm that the non-trivial data of the Lunin-Mathur geometry
are encoded in $K^3$ and $M$ from the relation \eqref{jtou17Feb21}.  The
duality equation between $A$ and $B$ (the second equation in
\eqref{def_ALM}) means that $d(-\ALM+\BLM)=dB^{3}$ is the
self-dual part of $d(-\ALM)=dk$, namely $(1+*_4)dk=dB^{3}$.  From
this, we can derive
\begin{subequations} 
\label{hlbt8Apr21}
 \begin{align}
 \mu *_3 dA+d\omega - V*_3 d\mu &= V^{-1}K^3 dA+d\xi^3,\label{mgvh17Feb21}\\
  -d\mu + \mu V^{-1}dV+V^{-1}*_3 d\omega &= -d(V^{-1}K^3).\label{mhlj17Feb21}
 \end{align}
\end{subequations}
Here we used the relations between $*_4$ and $*_3$ on the Gibbons-Hawking space \eqref{GHmetric},
\begin{align}
 \begin{gathered}
 *_4 \lambda _{(2)}=V^{-1}(d\psi+A)\wedge *_3 \lambda_{(2)} ,\qquad
 *_4 ((d\psi+A)\wedge \theta_{(1)})=V {*_3 \theta_{(1)}},
 \\
 *_4 \theta_{(1)} = (d\psi +A)\wedge *_3 \theta_{(1)},\qquad
 *_4(d\psi+A)=V^2 d^3y,
 \end{gathered}
\end{align}
where $\lambda_{(2)}$ is a 2-form and $\theta_{(1)}$ is a 1-form in $\bbR^3$.
The two equations in \eqref{hlbt8Apr21}
 are $*_3$ of each other.
By acting with $d$ on \eqref{mgvh17Feb21} we can derive
 \begin{gather}
  d*_3dM=0, \qquad
 \mu = M+{1\over 2}V^{-1}K^3.\label{fkpp18Feb21}
 \end{gather}
Plugging  \eqref{fkpp18Feb21} into \eqref{mhlj17Feb21}, we find
\begin{align}
 *_3 d\omega = V dM-M dV-{1\over 2}dK^3.
\label{edto19Mar21}
\end{align}
Equations \eqref{fkpp18Feb21} and \eqref{edto19Mar21} are consistent
with the relations \eqref{hljz8Apr21} and \eqref{sdomega} under the
identification
\begin{align}
 V={1\over r},\qquad L_1=Z_1,\qquad L_2=Z_2,\qquad L_3=1,\qquad K^1=K^2=0.
\label{jefl9Apr21}
\end{align}
So, the non-trivial data of the solution will be in $K^3$ and $M$.

\bigskip
Using the above dictionary \eqref{jtou17Feb21}, we can readily check
that the Lunin-Mathur solution based on the circular
profile~\eqref{profile_circular} corresponds to a harmonic solution with
the harmonic functions given in \eqref{VKLM_supertube} in the
introduction.

\section{Codimension-2 Lunin-Mathur solution}
\label{sec:codim-2_LM}

Here we explicitly construct harmonic functions that describe a
Lunin-Mathur geometry with both dipole charges in \eqref{D1+D5->KKM2}
(or equivalently \eqref{D2+D2->D4+NS5}).  Specifically, we consider the
following profile:
\begin{align}
 g_1+ig_2=a e^{i k\Omega \lambda},\qquad
 g_3+ig_4=b e^{-i k' \Omega \lambda}
\label{profile_helix}
\end{align}
with $a,b\ge 0$\,\footnote{We could make $a,b$ complex but that does not
make a difference after smearing.} and $k,k'\in\bbZ_{>0}$.\footnote{The
negative sign in front of $k'$ in $g_3+ig_4$ is because of the holographic
dictionary \cite{Giusto:2019qig}; see section~\ref{sec:cft}.
  } We have already discussed this
profile in \eqref{profile_helix0} in the introduction.  When smeared
along and reduced on $\psi$, this gives a circular ring in the
base~$\bbR^3$:
\begin{align}
 \textstyle\rho\equiv \sqrt{y_1^2+y_2^2}=R,\qquad y_3=c,\label{circle_in_R3}
\end{align}
where
\begin{align}
 R={ab\over 2},\qquad c={-a^2+b^2\over 4}.\label{Rc_ab}
\end{align}
This is the codimension-2 solution that we would like to construct and
study. 

Of course, we can derive the harmonic functions ``top-down'' by starting
with the Lunin-Mathur geometry in 6d with the profile
\eqref{profile_helix}, smearing and reducing it to 4d/5d, and then reading
off the harmonic function.  However, here we go ``bottom-up'' by
directly constructing the harmonic functions based on the expected
charges that they must represent.  In the next section, we will confirm
the result from the top-down viewpoint starting from the Lunin-Mathur
geometry.

\bigskip
The profile \eqref{profile_helix}
is going in the angular directions
$\phi,\psi$ as
\begin{align}
 \phi=-(k+k')\Omega \lambda,\qquad
 \psi=2k\Omega\lambda.
\label{phipsi_lambda}
\end{align}
Therefore, by the argument below \eqref{D1+D5->KKM2} we will have $-(k+k')$
units of NS5 charges and $k$ units of D4(4567) charges, encoded in
harmonic functions.

As mentioned in the introduction, $a$ and $b$ are constrained to satisfy
\eqref{ghfp19Apr21} (this will be shown in \eqref{jsjr14Apr21}). If we
fix the charges of the system, $Q_1$ and $Q_5$, and also the parameters
$k$ and $k'$, then the ring will be on a spheroid in $\bbR^3$  for any
values of $a$ and $b$.  More precisely, the position of the ring
$\rho=R=ab/2$, $y_3=c=(-a^2+b^2)/4$  can be shown to satisfy
\begin{align}
 \biggl({y_3-y_3^{(0)}\over A}\biggr)^2
 +
 \biggl({\rho\over B}\biggr)^2
=1,
\label{spheroid}
\end{align}
where
\begin{align}
 A\equiv {\cN\over 8}\left({1\over k^2}+{1\over k'^2}\right),\quad
 B\equiv {\cN\over 4kk'},\quad
 y_3^{(0)}\equiv {\cN\over 8}\left({1\over k'^2}-{1\over k^2}\right),\quad
\cN\equiv {Q_1\over Q_5\Omega^2}={Q_1 Q_5\over R_y^2}.
 \label{kwwp20Apr21}
\end{align}
The surface
\eqref{spheroid}
 is a spheroid, whose symmetry axis is
the $y_3$ axis.  The origin $\yv=0$ is a focal point of the ellipse on the
cross section that contains the $y_3$ axis.  Because $A\ge B$,
the spheroid is prolate.

We will be interested in the process of starting with $a>0,b=0$ and then
increasing $b$, finally ending with $a=0,b>0$.
When $a>b$, the $y_3$ coordinate of the ring is $c<0$.  We call this the
``southern'' case, because the ring is like a latitude line on the
southern hemisphere of the spheroid.  When $a<b$, the $y_3$ coordinate
of the ring is $c>0$.  We call this the ``northern'' case.  However, the
word ``southern''/``northern'' should not be taken literally, because
the center of the spheroid is not at $y_3=0$.
See Figure
\ref{fig:spheroid0}.

As the extreme cases, if $a>0,b=0$ (the ``south pole'' limit), the 4d
profile is a circle of radius $a$ on the $x_1$-$x_2$ plane discussed in
\eqref{profile_circular}, which projects in $\bbR^3$ onto a point on the
negative $y_3$ axis ($y_3=-a^2/4$).  In this case, the $\phi$ circle
shrinks and there is no NS5 charge. So, this limit gives a codimension-3
solution.  If $a=0,b>0$ (the ``north pole'' limit), on the other hand,
the profile is a circle of radius $b$ on the $x_3$-$x_4$ plane, which
projects onto a point on the positive $y_3$ axis ($y_3=b^2/4$).  So,
this limit also gives a codimension-3 solution. In this case, the
shrinking cycle is not $\phi$ or $\psi$ but their linear combination
because of the nontrivial Hopf fibration.

\subsection{Building blocks}

To find the harmonic solution that corresponds to the profile
\eqref{profile_helix}, we start by constructing appropriately normalized
harmonic functions that have codimension-2 and codimension-3 sources
along the ring \eqref{circle_in_R3}, as building blocks.

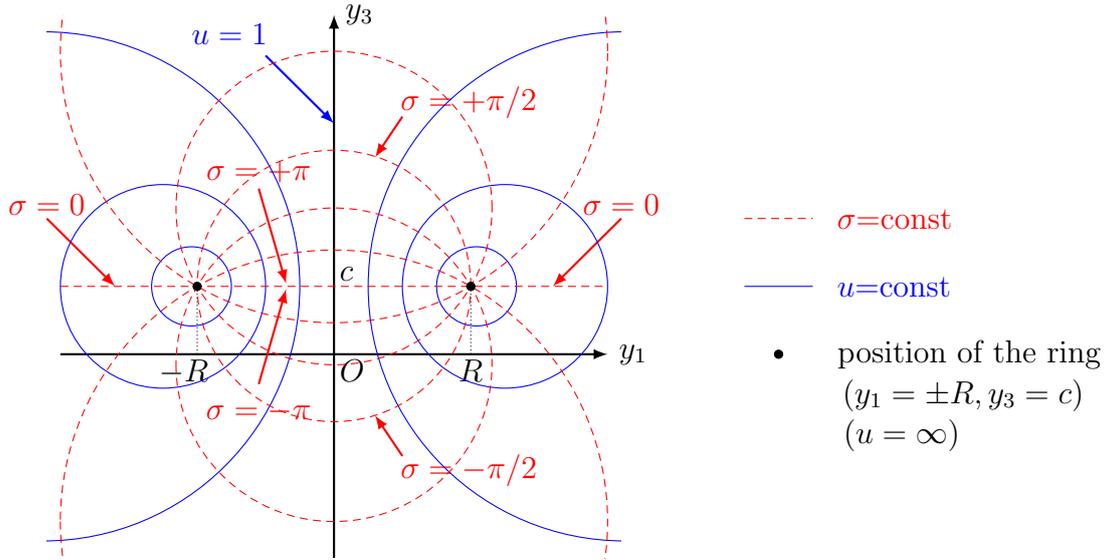
\begin{figure}[htb]
\begin{center}
  \begin{tikzpicture}[scale=1.8]
 \def\yoffset{1/2}
 \draw[densely dashed,color=red] (3.0,1) -- +(.5,0) node [right,xshift=5] {$\sigma$=const};
 \draw[color=blue] (3,0.5) -- +(.5,0) node [right,xshift=5] {$u$=const};

 \draw[fill=black] (3.25,0) circle (0.03) node [right,xshift=18] {position of the ring};
 \draw (4.6,-0.3) node {$(y_1=\pm R,y_3=c)$};
 \draw (4.15,-0.6) node {$(u=\infty)$};

 \draw[-latex,thick] (-2,0) -- (2,0) node [right] {$y_1$};
 \draw[-latex,thick] (0,-1.5) -- (0,2.5) node [right] {$y_3$};

 \draw[thick,latex-,color=blue] (0,1.7) -- +(-0.5,0.5) node [above left,xshift=5,yshift=0] {$u=1$};

 \draw[thick,color=red,latex-] ( 1.6,\yoffset) -- +( 0.5,0.5) node [above,yshift=-3] {$\sigma=0$};
 \draw[thick,color=red,latex-] (-1.6,\yoffset) -- +(-0.5,0.5) node [above,yshift=-3] {$\sigma=0$};

 \draw[thick,color=red,latex-] (-0.35,\yoffset+0.02) -- +(-0.2, 0.7) node [above,yshift=-2] {$\sigma=+\pi$};
 \draw[thick,color=red,latex-] (-0.35,\yoffset-0.02) -- +(-0.2,-0.7) node [below,yshift=-2] {$\sigma=-\pi$};

 \draw[thick,color=red,latex-] ( 0.30,\yoffset+.95) -- +(0.2, 0.3) node [above right,yshift=-5,xshift=-5] {$\sigma=+\pi/2$};
 \draw[thick,color=red,latex-] ( 0.30,\yoffset-.95) -- +(0.2,-0.3) node [below right,yshift= 5,xshift=-5] {$\sigma=-\pi/2$};

 \draw (0,\yoffset) node [above right,yshift=-2,xshift=-2] {$c$};

 \draw (0,0) node [below right,yshift=2,xshift=-2] {$O$};

 \draw[densely dotted] ( 1,\yoffset) -- ( 1,0) node [below,yshift=2] {$R$};
 \draw[densely dotted] (-1,\yoffset) -- (-1,0) node [below,yshift=2,xshift=-5] {$-R$};

 \clip (-2.1,-1.5) rectangle (2.1,2.5); 

 \foreach \u in {17/15, 5/3, 25/7} {
 \draw[color=blue] ({\u/sqrt(\u*\u-1)},\yoffset) circle ({1/sqrt(\u*\u-1)});
 \draw[color=blue] ({-\u/sqrt(\u*\u-1)},\yoffset) circle ({1/sqrt(\u*\u-1)});
 } 

   \foreach \sigma in {1/6,1/3,0.499,-1/3,-1/6} {
 \draw[densely dashed,color=red] (0,{1/tan(deg(\sigma*pi))+\yoffset}) circle ({1/sin(deg(\sigma*pi))});
 } 
 \draw[color=red,densely dashed] (-2,\yoffset) -- (2,\yoffset);

 \draw[fill=black] ( 1,\yoffset)circle (0.03);
 \draw[fill=black] (-1,\yoffset)circle (0.03);
 \end{tikzpicture}
 \caption{Toroidal coordinates (on the $y_2=0$ plane).\label{fig:toroial_coords}}
\end{center}
\end{figure}

It is
useful to introduce  toroidal coordinates $(u,\sigma,\phi)$, adapted
for a ring sitting at $\sqrt{y_1^2+y_2^2}=R$, $y_3=c$:
\begin{align}
 y_1={\sqrt{u^2-1}\over u-\cos\sigma}R\cos\phi,\qquad
 y_2={\sqrt{u^2-1}\over u-\cos\sigma}R\sin\phi,\qquad
 y_3'\equiv y_3-c={\sin\sigma\over u-\cos\sigma}R,
\label{hdtf19Mar21}
\end{align}
where $1\le u<\infty$, $\phi\cong\phi+2\pi$, $\sigma\cong\sigma+2\pi$.
The ring sits at $u=\infty$, while $u=1,\sigma=0$
corresponds to $r=|\yv|\to\infty$. $\phi$ is the angle along the ring
while $\sigma$ is the angle going around the ring.  See Figure~\ref{fig:toroial_coords} for a graphical description.
The flat 3d metric
can be written as
\begin{align}
 d\yv^2&=dy_1^2+dy_2^2+dy_3'{}^2={R^2\over (u-\cos\sigma)^2}
 \left[{du^2\over u^2-1}+{d\sigma^2}+(u^2-1)d\phi^2\right].
\end{align}
The inverse relations are 
\begin{align}
 \cos\sigma={\yv'^2-R^2\over \Lambda},\qquad
 u={\yv'^2+R^2\over \Lambda},\qquad
 \Lambda^2=(\yv'^2-R^2)^2+4R^2 y_3'{}^2,
\label{ghfy24Feb21}
\end{align}
where $\yv'{}^2=y_1^2+y_2^2+y_3'{}^2$.

Near the ring, $u\to \infty$,
\begin{align}
 {\textstyle\sqrt{y_1^2+y_2^2}}\approx R+\varrho \cos\sigma,\quad
 y_3'=y_3-c\approx \varrho \sin\sigma,
\qquad
 \varrho \equiv {R\over u}.
\end{align}
Namely, $(\varrho,\sigma)$ are plane polar coordinates near the position of
the ring.

\medskip
Let us first introduce $H$, which is a single-valued harmonic function
with codimension-3 sources uniformly distributed along the ring.
Explicitly, it is given by
\begin{align}
 H&={1\over R}\sqrt{{u-\cos\sigma\over 2}}\, P_{-1/2}(u)
 =
{4\,\cI_{00}(P,Q),\over s^2+a^2+w^2+b^2}
\end{align}
where $P_n(u)$ is the Legendre function and we defined the following
quantities:
\begin{subequations} 
 \begin{gather}
 P\equiv {2sa\over s^2+a^2+w^2+b^2},\qquad
 Q\equiv {2wb\over s^2+a^2+w^2+b^2},
\label{def_PQ}
 \\
 \cI_{mn}(P,Q)\equiv 
 \int_0^{2\pi}{d\mu\over 2\pi}
 \int_0^{2\pi}{d\nu\over 2\pi}\,
 {(P\cos\mu)^m (Q\cos\nu)^n\over 1-P\cos\mu - Q\cos\nu}.
\label{def_Imn}
 \end{gather}
\end{subequations}
For other expressions of $H$, see Appendix \ref{app:funcs}\@. The behavior of $H$ near the ring and near
infinity is
\begin{align}
 H
 =
\begin{cases}
 \displaystyle {\log(8u)\over \pi R}+\cO\Bigl({1\over u}\Bigr)  & \text{(near the ring, $u\to \infty$),}\\[2ex]
 \displaystyle {1\over r}+\cO\Bigl({1\over r^2}\Bigr)  & \text{(near infinity, $r\to\infty $).}\\
\end{cases}
\label{nfha8Apr21}
\end{align}
Because $H$ is harmonic, $\bigtriangleup H$ vanishes except on the
ring. The first line of \eqref{nfha8Apr21} implies that $\bigtriangleup
H$ has delta-function singularities uniformly smeared along the ring,
and the total charge, as measured by the coefficient of $1/r$ for large
$r$, is unity.

Next, we want a harmonic function---call it $\gamma$---which has the
following monodromy around the ring:
\begin{align}
 \gamma(u,\sigma+2\pi)\to \gamma(u,\sigma)+1.
\label{jxwa11Apr21}
\end{align}
Actually, let us be more general for a bit and consider a harmonic function
$\gamma$ with the monodromy $\gamma\to \gamma+1$ around a general closed curve
$\cC$ in $\bbR^3$.  For that, it is convenient to introduce a 1-form
\begin{align}
 \alpha = dy_i \int_0^{2\pi}{d\mu\over 4\pi} 
 {\p_\mu f_i(\mu)\over |\yv-\fv(\mu)|},\label{njrj11Apr21}
\end{align}
where $f_i(\mu)$, $0\le\mu\le 2\pi$, is a 3d profile function that
parametrize the curve $\cC$ (this $\alpha$ is a 3d version of $\ALM$ defined in \eqref{def_ALM}). The 1-form $\alpha$ is independent of
how we parametrize $\fv(\mu)$. It is not difficult to show that, if we
define $\gamma$ by the condition
\begin{align}
 d\gamma=*_3d\alpha,\label{mvbe11Apr21}
\end{align}
then $\gamma$ is harmonic and has the desired monodromy $\gamma\to
\gamma+1$ as we go around $\cC$.  One can also show \cite{Park:2015gka}
that $\bigtriangleup \gamma$, which could have delta-function
singularities on $\cC$, is identically zero:
\begin{align}
 \bigtriangleup \gamma=0\qquad \text{(no delta function)}.\label{Lap_gamma==0}
\end{align}
This will be important when we impose the integrability condition.

In the present case, curve $\cC$ is the circular ring
\eqref{circle_in_R3}.  Being careful to the orientation of the angle
variable $\sigma$ relative to the ring (see Figure~\ref{fig:toroial_coords}), we take the 3d profile function to be
\begin{align}
 f_1+if_2=Re^{-i\mu},\qquad f_3=c.\label{njrp11Apr21}
\end{align}
The 1-form $\alpha$ for this profile is found to be \cite{deBoer:2012ma}
\begin{align}
 \alpha(u,\sigma)&={a(u)\over\sqrt{u-\cos\sigma}}d\phi,\qquad
 a(u)=
-{R\over 8\sqrt{2}}
 {u^2-1\over u^{3/2}}\,\,_2F_1\left(\tfrac{3}{4},\tfrac{5}{4};2; 1-u^{-2}\right).
\label{hlst2Feb17}
\end{align}
The function  $\gamma$ that satisfies \eqref{mvbe11Apr21} is
\begin{align}
 \gamma(u,\sigma)&={1\over R}
\left[
 {a(u)\over\sqrt{u-1}}\,{\bf F}(\tfrac{\sigma}{2}|-\tfrac{2}{u-1})
 -2\sqrt{u-1}\,a'(u)\,{\bf E}(\tfrac{\sigma}{2}|-\tfrac{2}{u-1})
 \right],
\label{egzm3Mar21}
\end{align}
where ${\bf F}(\phi|m)$ and ${\bf E}(\phi|m)$ are the elliptic integrals
of the first and second kinds, respectively.\footnote{We follow the
Mathematica convention for the arguments of elliptic integrals.}  One
can check the monodromy \eqref{jxwa11Apr21} from the periodicity of the
elliptic integrals.  This $\gamma$ vanishes at infinity
$r=|\yv|\to\infty$, which corresponds to $\sigma= 0,u\to 1$ (and if we
go to other branches $\sigma= 2\pi n,u\to 1$ with $n\in\bbZ$ then $\gamma\to n$; a related formula is \eqref{gamma(u=1)}).
The near-ring behavior of $\gamma$ is found to be
\begin{align}
 \gamma(u,\sigma)={\sigma\over 2\pi}+\cO\Bigl({1\over u}\Bigr).
\end{align}

\subsection{Codimension-2 source}
Using $H,\gamma$ defined above, let us find harmonic functions that
represent the Lunin-Mathur geometry with the profile
\eqref{profile_helix}.  First, we have
\begin{align}
 V={1\over r},\qquad
 Z_1=L_1={Q_1\over 4}H,\quad
 Z_2=L_2={Q_5\over 4}H,\qquad
 L_3=1,\qquad K^1=K^2=0.
\label{jjyn16Apr21}
\end{align}
$V$ is always the same for the Hopf fibration, while $Z_{1,2}$ are
determined so that the total charge is the same as in the south pole
limit 
\eqref{VKLM_supertube}
(or 
\eqref{Z1Z2_south_pole}) but now the charges must distributed
uniformly along the ring.  The other harmonic functions are fixed by
\eqref{jefl9Apr21}.

Before studying what singularities $K^3$ must have, let us discuss its
normalization.  When there are $n$ units of KKM($6789\psi,y$) charge at
$r=0$, $K^3$ has the following codimension-3 center:
\begin{align}
 K^3\supset{R_y n\over 2r}
={Q_5 \Omega n\over 2r},
\end{align}
where we used \eqref{def_Omega}.  On the
other hand, when there are $n'$ units of KKM($6789\cC,y$) charge lying
along curve $\cC$, then $V^{-1}K^3$ has the following monodromy as we go
around $\cC$:
\begin{align}
 {\mathit\Delta}(V^{-1}K^3)
\equiv
 V^{-1}K^3|_{\sigma+2\pi}
 - V^{-1}K^3|_{\sigma} 
 ={R_y n'\over R_\psi}
 ={Q_5 \Omega n'\over 2}
\label{Delta(K3/V)}
\end{align}
where $R_\psi=2$.  These normalizations can be derived by standard
arguments (see e.g.~\cite{Peet:2000hn}).

From \eqref{phipsi_lambda}, we know that we have $-(k+k')$ units of
KKM($6789\cC,y$) lying along $\phi$.  
So, $K^3$ must have the monodromy 
\eqref{Delta(K3/V)} with $n'=-(k+k')$.
Clearly, this requirement is satisfied if we take $K^3$ to be:
\begin{align}
 K^3 =-{Q_5 \Omega (k+k')\over 2}\,(\gamma V +\Kt^3),\label{gyql21Apr21}
\end{align}
where $\Kt^3$ is a single-valued function.  For $K^3$ to be harmonic,
$\Kt^3$ must satisfy
\begin{align}
 \bigtriangleup \Kt^3
 = -{\bigtriangleup(\gamma V)}
= -\p_i \gamma\, \p_i V
 = -\epsilon_{ijk}\p_j \alpha_k\, \p_i V,
\end{align}
where in the second equality we used that $\gamma,V$ are harmonic and in
the last equality we used~\eqref{mvbe11Apr21}.  A special solution of
this Poisson equation is
\begin{align}
 \Kt^3=-\int_0^{2\pi} {d\mu\over 4\pi}\,
 {\epsilon_{ijk}\, y_i f_j\, \p_\mu f_k\over r\,|\fv|\,|\yv-\fv|\,(r+|\fv|+|\yv-\fv|)},
\label{jijj25Mar21}
\end{align}
where $\alpha$ and $\fv$ are given in \eqref{njrj11Apr21} and
\eqref{njrp11Apr21}.  In the present case where $f_i$ is given by
\eqref{njrp11Apr21}, this can be evaluated as
\begin{align}
 \Kt^3
 &=-{R\over r |\fv|}
 \int_0^{2\pi} {d\mu\over 4\pi}\,
 {c\rho \cos(\mu+\phi) - R y_3\over
 |\yv-\fv|\,(|\fv|+|\yv-\fv|+r)} 
\notag\\
 &=-
 {4ab\,V\over (a^2+b^2)(s^2+w^2+a^2+b^2)^2}
 \int_0^{2\pi} {d\mu'\over 4\pi}\,
 {(b^2-a^2)sw\cos\mu' + ab(s^2-w^2)\over 
 \sqrt{X}(1+\sqrt{X})
 }\,,
\label{euto20Apr21}
\end{align}
where $\mu'=\mu+\phi$, $ X\equiv 1-(P^2+Q^2)-2PQ\cos\mu'$ ($P,Q$ were defined in
\eqref{def_PQ}).  In the first equality we went to cylindrical
coordinates $(\rho,\phi,y_3)$ in $\bbR^3$ and in the second equality we used
relations such as \eqref{fsfz13Apr21}, \eqref{fsgo13Apr21}, and \eqref{fsgx13Apr21}.
  One can
show 
(see Appendix \ref{app:various_rel})
 that this can be written in terms of $\cI_{mn}(P,Q)$ as
\begin{align}
\Kt^3&=
 -{b^2\cI_{10}(P,Q)-a^2 \cI_{01}(P,Q)\over a^2+b^2}V
\notag\\
 &=
 -{1\over a^2+b^2}\,V
 \int_0^{2\pi}{d\mu\over 2\pi}
 \int_0^{2\pi}{d\nu\over 2\pi}\,
 {b^2 P\cos\mu - a^2Q\cos\nu\over 1-P\cos\mu - Q \cos\nu}
 \, .
\label{eumb20Apr21}
\end{align}
So, we have fixed the form of $K^3$ with the desired codimension-2
source to be
\begin{align}
 K^3 =-{Q_5 \Omega \over 2}(k+k')\left(\gamma 
 -{b^2\cI_{10}-a^2 \cI_{01}\over a^2+b^2}\right)V
+K^3_{(3)}
\label{lloa12Apr21}
\end{align}
where $K^{3}_{(3)}$ is a single-valued harmonic function containing only
codimension-3 sources, which we turn to next.  $\cI_{10},\cI_{01}$ can
be expressed in terms of elliptic integrals; see Appendix \ref{app:funcs}.

\subsection{Codimension-3 sources}

We want $K^3$ to also have $k$ units of codimension-3 D4-brane charge
along the ring, as discussed in \eqref{phipsi_lambda}.  We cannot simply
set $K^3_{(3)}$ to $kH$, because the first term of
\eqref{lloa12Apr21} also already contains some codimension-3 charge,
which we must take into account.

Codimension-3 sources in a harmonic function will appear as
$\delta$-function singularities when we act with $\bigtriangleup$ on it.  So,
let us examine the expression
\begin{align}
\bigtriangleup \left[\left(\gamma 
 -{b^2\cI_{10}-a^2 \cI_{01}\over a^2+b^2}\right)V\right].
\end{align}
From the first term, we have $\bigtriangleup (\gamma
V)=(\bigtriangleup\gamma)V+2\partial_i\gamma\,\partial_i
V+\gamma(\bigtriangleup V)$ but $(\bigtriangleup \gamma) V=0$
from~\eqref{Lap_gamma==0} while the cross term
$\partial_i\gamma\,\partial_i V$ contains no $\delta$ function.  So,
only $\gamma(\bigtriangleup V)$ remains.  For the second term,
$[\bigtriangleup (b^2 \cI_{10}-a^2 \cI_{01})]V$ has no $\delta$ function
from \eqref{mvhq12Apr21}, while $(b^2 \cI_{10}-a^2
\cI_{01})\,{\bigtriangleup V}$ vanishes by \eqref{pty20Apr21}. There
is no $\delta$ function from the cross term.  So, after all, the only
$\delta$-function singularities representing localized codimension-3
charges are
\begin{align}
\bigtriangleup K^3 
 &=-{Q_5 \Omega \over 2}(k+k')\,\gamma(r=0)\, {\bigtriangleup V}
+\bigtriangleup K^3_{(3)}
\notag\\
 &=
\begin{cases}
  \displaystyle
 -{Q_5 \Omega \over 2}(k+k'){b^2\over a^2+b^2}\, {\bigtriangleup V}+\bigtriangleup K^3_{(3)}  &\qquad \text{($a>b$, ``southern'')}, \\[2ex]
  \displaystyle
 +{Q_5 \Omega \over 2}(k+k'){a^2\over a^2+b^2}\, {\bigtriangleup V}+\bigtriangleup K^3_{(3)}  &\qquad \text{($b>a$, ``northern'')}, \\
\end{cases}
\label{emu17Apr21}
\end{align}
where in the second equality we used \eqref{fsci21Mar21} for the value
of $\gamma(r=0)$. So, we have $-(k+k'){b^2 \over a^2+b^2}$ units of D4
charge sitting at $r=0$ in the ``southern'' case, and $(k+k'){a^2 \over
a^2+b^2}$ units for the ``northern'' case.  
 The discontinuity as we go from the south to the north
will be discussed in section \ref{ss:comments}.

\subsubsection*{Delocalized charge}

Actually, this is not the end of the story.  In the current situation
where we have a branch cut of the multi-valued function $\gamma$, we can
also have delocalized charges.

\begin{figure}[htb]
\begin{center}
 \begin{tikzpicture}[scale=0.7]
  \draw[-latex] (-3,0) -- (3,0) node [right] {$y_1$};
  \draw[-latex] (0,-3) -- (0,3) node [left] {$y_3$};
  \fill[black!50!green] (0,0) circle (0.1);
  \draw[red,thick,decorate,decoration={zigzag,amplitude=1,segment length=4}] (-1,-1.5) -- (1,-1.5);
  \fill[blue] (-1,-1.5) circle (0.1);
  \fill[blue] (1,-1.5) circle (0.1);

  \draw[thick,black!40!white] (0,0) circle (2.5);
  \draw[black!40!white] (2.2,2.2) node {$S_\infty^2$};

  \draw[thick,black!40!white] (0,0) circle (.2) node [above right] {$S^2_0$};

  \draw[thick,black!40!white] (-0.84,-1.6) -- (0.84,-1.6);
  \draw[thick,black!40!white] (-0.84,-1.4) -- (0.84,-1.4);

  \draw [thick,black!40!white,domain=30:330] plot ({-1+0.2*cos(\x)}, {-1.5+0.2*sin(\x)});
  \draw [thick,black!40!white,domain=-150:150] plot ({1+0.2*cos(\x)}, {-1.5+0.2*sin(\x)}) node 
   [above right] {$S^2_1$};
  \draw[black!40!white] (0,-1.1) node {``in''};
  \draw[black!40!white] (0,-1.9) node {``out''};
 \end{tikzpicture}
\begin{quote}
  \caption{\sl The Gaussian surfaces. The large gray circle represents
 $S^2$ at infinity, which can be deformed into two disconnected
 surfaces: $S^2_0$, enclosing the origin, and $S^2_1$, surrounding the
 disk $D_2$ (red zigzag line) whose boundary is the ring (blue
 dots).  The difference of a quantity $X$ across $D_2$ is defined to be
 $[X]=X_{\rm out}-X_{\rm in}$, where ``out'' and ``in'' are defined as
 shown here, for the southern case.  \label{fig:gauss_sfc_1}}
\end{quote}
\end{center}
\end{figure}
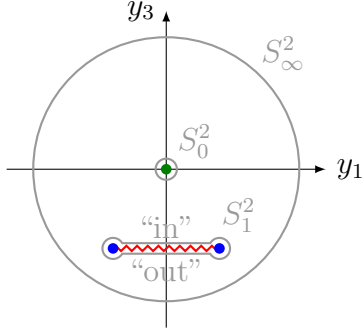

Recall that the D4 charge is measured by the Gaussian integral inside
$\bbR^3$ as
\begin{align}
q_{\rm D4}^{}= -{1\over 4\pi}\int_{S^2_\infty} *_3 dK^3,
\label{gnxo21Mar21}
\end{align}
where $S^2_\infty$ is the sphere at infinity, with its normal pointing
outward. We can continuously deform $S^2_\infty$ into two disconnected
pieces one of which encloses $r=0$ and the other of which surrounds the
disk $D_2$ whose boundary is the ring (see Figure~\ref{fig:gauss_sfc_1}).  Because $K^3$ is discontinuous across the disk,
there will be non-trivial contribution to the integral there.  The only
discontinuous part in $dK^3$ is
\begin{align}
  dK^3|_{\rm discont}
 =-{Q_5 \Omega \over 2}(k+k')
 d(\gamma V)|_{\rm discont}
 =-{Q_5\Omega \over 2} (k+k')
\gamma\,  dV
\end{align}
where we used the fact that $d\gamma=*_3 d\alpha$ is continuous.  Let us
denote the discontinuity of $X$ by $[X]=X_{\rm out}-X_{\rm in}$, where
``out'' and ``in'' mean outside and inside of $D_2$, with the
orientation derived from that of $S^2_\infty$, as shown in Figure~\ref{fig:gauss_sfc_1}, in the ``southern'' case ($a>b$).
In the ``southern''  case, we have  $[\gamma]=-1$ and 
the contribution to \eqref{gnxo21Mar21} from the delocalized charge is
\begin{align}
q_{\rm deloc}^{}
 &= -{1\over 4\pi}\int_{D_2} [*_3 dK^3]
 ={1\over 4\pi}
 {Q_5\Omega \over 2}(k+k')
\int_{D_2} [\gamma]*_3 dV\notag\\
 &=-{1\over 4\pi}
 {Q_5\Omega \over 2}(k+k')
\int_{D_2} *_3 dV
 =-{1\over 4\pi}
 {Q_5\Omega \over 2}(k+k')
\int_{D_2} \sin\theta \, d\theta\wedge d\phi.
\end{align}
In the ``southern'' case ($a>b$), the $\theta$ integral is from $\theta$
such that $\cos\theta={c\over \sqrt{c^2+R^2}}={b^2-a^2\over a^2+b^2}$ to
$\theta=\pi$.  So, evaluating the integral, we find
\begin{align}
 q_{\rm deloc}^{}
 &=
 \begin{cases}
  \displaystyle
 +{Q_5\Omega \over 2}(k+k')
{b^2\over a^2+b^2}
  &\qquad \text{($a>b$, ``southern'')}\\[2ex]
  \displaystyle
 -{Q_5\Omega \over 2}(k+k')
{a^2\over a^2+b^2}
  &\qquad \text{($b>a$, ``northern'')}\\
 \end{cases}
\end{align}
We also included the expression for the ``northern'' case (in which
$[\gamma]=+1$).  The discontinuity as we go from the south to the north
will be discussed in section \ref{ss:comments}.
The sum of this delocalized charge and the
charge localized at $r=0$ in \eqref{emu17Apr21} is zero.

\subsubsection*{The expression for $K^3$}

The codimension-3 part $K^3_{(3)}$ is chosen so that the ring carries
$k$ units of D4 charge (including the localized and delocalized charges)
in the southern case:
\begin{align}
 K^3_{(3)}={Q_5\Omega\over 2}{a^2k-b^2k'\over a^2+b^2}(H-V).
\end{align}
Actually, here we added a term proportional to $V$, which can be thought
of as the ``gauge transformation'' \eqref{gauge_transf}.  We will see
that this choice agrees with the 6d result in the next section.
For continuity, we must take the same $K^3_{(3)}$ also for
the northern case.

So, the final expression for $K^3$ is
\begin{align}
 K^3 ={Q_5 \Omega \over 2}\left[-(k+k')\left(\gamma 
 -{b^2\cI_{10}-a^2 \cI_{01}\over a^2+b^2}\right)V
 +{a^2k-b^2k'\over a^2+b^2}(H-V)\right].
\label{K3_final}
\end{align}

\subsection{Integrability  and no-CTC conditions}

Having determined $K^3$, we can find the remaining harmonic function
$M$.  First, It must contain $\gamma$ so that $\mu$ \eqref{fkpp18Feb21}
is single-valued.  Furthermore, it must satisfy the integrability
condition derived from \eqref{edto19Mar21},
\begin{align}
 0=V{\bigtriangleup M}-M{\bigtriangleup V}-{1\over 2}{\bigtriangleup K^3}
 \qquad \text{(no $\delta$ function).}
\end{align}
Imposing these conditions, it is straightforward to show that $M$ is given by:
\begin{align}
 M={Q_5 \Omega \over 4}\left[
 (k+k')\gamma + {1\over 4}(ka^2-k'b^2)H \right].
\label{M_final}
\end{align}
The 1-form $\omega$ that satisfies the equation \eqref{edto19Mar21} is
\begin{align}
 \omega=-{Q_5\Omega\over 2}
 \left(
 k\cI_{10}\cos^2{\theta\over 2}
 +k'\cI_{01}\sin^2{\theta\over 2}
 \right)d\phi.
\end{align}
The modulus $\tau^3$ (see
\eqref{def_tau^I}) is found to be
\begin{align}
 \tau^3=-(k+k')\gamma+\text{(single-valued)},
\end{align}
which has the monodromy $\tau^3\to \tau^3-(k+k')$ due to the $-(k+k')$
NS5($4567\lambda$) branes.  This monodromy can be represented by an
$SL(2,\bbZ)$ transformation
\begin{align}
\cM = \begin{pmatrix}
  1&-(k+k') \\
 0 & 1
 \end{pmatrix}.\label{fenv18May21}
\end{align}

\bigskip At this point, the solution contains $Q_1,Q_5,a,b,k,k'$ as
independent parameters.  In the Lunin-Mathur geometry, they are going to
be related by the relation \eqref{def_Q1}; for example, in the
south-pole limit it is given by \eqref{Q1_south_pole}.  In the current
framework of harmonic solutions in 4d/5d, such a relation can be derived by
studying the no-CTC condition.

Let us focus on the 5d part of the 11d metric \eqref{Msol}.
If we focus on the $\phi,\psi$ part, we have
\begin{align}
 Z^{2/3}ds^2_5
 \supset
 -[\mu(d\psi+A_\phi d\phi)+\omega_\phi d\phi]^2
 +Z[V^{-1}(d\psi+A_\phi d\phi)^2+Vr^2\sin^2\theta\,d\phi^2]
\label{gizv14Apr21}
\end{align}
where we used \eqref{k1form}.  Near the ring, $u\sim \infty$, the
quantities appearing here behave as (see \eqref{pvw20Apr21},
\eqref{pty20Apr21})
\begin{align}
 \begin{gathered}
 \omega_\phi
 \sim {Q_5\Omega\over 2}
 {ab\over a^2+b^2}(k+k'){\log u\over \pi},\quad
 \mu 
 \sim {Q_5\Omega\over 4} \left(
 k{a\over b}-k'{b\over a}
 \right){\log u\over \pi},
 \\
 A_\phi \sim {2b^2\over a^2+b^2},\quad
 V\sim {4\over a^2+b^2},\quad
 Z\sim {Q_1 Q_5\over 4\pi^2a^2b^2}\log^2u,\quad
 g_{\phi\phi}=
 r^2\sin^2\theta
 \sim {a^2b^2\over 4}.
 \end{gathered}
\end{align}
By diagonalizing the metric \eqref{gizv14Apr21} and requiring that there
is no CTC, namely the eigenvalues are non-negative, we find the
condition
\begin{align}
 Q_1\ge \Omega^2(a^2k^2+b^2k'^2)Q_5.
\end{align}
In supertube configurations, genuine microstates saturate such non-CTC
inequalities \cite{Emparan:2001ux}.  Because Lunin-Mathur geometries are genuine
microstates, we must require
\begin{align}
 Q_1= \Omega^2(a^2k^2+b^2k'^2)Q_5.
\label{jsjr14Apr21}
\end{align}
This is what we used in \eqref{ghfp19Apr21}.

\subsection{Comments}
\label{ss:comments}

Thus far, in this section, we have demonstrated the phenomenon of charge
transfer, which is summarized in Figure~\ref{fig:spheroid0}.  The D4
charge has two parts, the one localized at the location of the ring, and
the other distributed over the branch cut inside the ring.  As we move
from the south pole to the north pole, the cut crosses the D6 center at
$r=0$ and, at that point, some of the D4 charge gets transferred to the
D6 center. As the result, the ring has different D4 charges in the
initial and final states.

This can be understood in terms of a ``gauge transformation''.  Let us
note that, if we apply the gauge transformation \eqref{gauge_transf}
with $c^1=c^2=0$, $c^3=-nQ_5(k'+k)\Omega/2$ to the harmonic functions
\eqref{K3_final}, \eqref{M_final}, we obtain
\begin{align}
\begin{split}
  K^3 &={Q_5 \Omega \over 2}\left[-(k+k')\left(\gamma +n
 -{b^2\cI_{10}-a^2 \cI_{01}\over a^2+b^2}\right)V
 +{a^2k-b^2k'\over a^2+b^2}(H-V) \right],\\
 M&={Q_5 \Omega \over 4}\left[
 (k+k')(\gamma+n) + {1\over 4}(ka^2-k'b^2)H  \right].
\end{split}
\end{align}

Let us consider the process of starting from the south pole limit with
$n=0$ and moving over to the north pole.  In the ``southern'' case
$a>b$, the branch cut is at $y_3=(b^2-a^2)/4<0$ and at the position of
the $r=0$ center, we have $0<\gamma<1/2$.  From \eqref{emu17Apr21}, we
see that the D4 charge of the $r=0$ center is $-{Q_5\Omega\over
2}(k+k'){b^2\over a^2+b^2}$.  If we increase $b$ and go over the to the
``northern'' case $a<b$, the $r=0$ pole has gone through the ring and
now $\gamma>1/2$ there.  If we want to bring it back to the
$-1/2<\gamma<1/2$ branch which can be connected to infinity (note that
$\gamma\to 0$ at infinity), we must do a gauge transformation with
$n=-1$ so that $-1/2<\gamma<1/2$ there.  This is how we get
${Q_5\Omega\over 2}(k+k'){a^2\over a^2+b^2}$ in \eqref{emu17Apr21} as
the D4 charge of the $r=0$ center. Namely, the ``jump'' in the D4 charge
happens because of the gauge transformation needed to bring the $r=0$
pole, which has gone to a different gauge by going through the ring,
back into the original gauge.

\bigskip
In the current note, we only discussed the ``helical'' profile
\eqref{profile_helix}, but we could consider any 4d profile and its
reduction to $\bbR^3$.  In $\bbR^3$, it will give an NS5-brane along
some arbitrary curve $\cC$, with D4 charge density which can vary along
$\cC$.  If we define a 1-form $\alpha$ as in \eqref{njrj11Apr21} (now by
an integration along $\cC$) and the dual scalar $\gamma$ by
$d\gamma=*_3d\alpha$, the construction of $K^3$ goes just as in
\eqref{gyql21Apr21}--\eqref{jijj25Mar21}, although 
explicit expressions will be harder to obtain.

\section{Deriving harmonic functions from 6d}
\label{sec:from_4+1}

In the previous section, we derived the harmonic functions with
codimension-2 sources that represent a Lunin-Mathur geometry in 6d,
based on the information about what charges must be present (bottom-up).
Here, we confirm that these harmonic functions are correct, starting
from 6d (top-down).

The various functions of the Lunin-Mathur geometry
\eqref{LM_geom_funcs} computed for the 
profile \eqref{profile_helix} are given by
\begin{align}
\begin{split}
  Z_1&={Q_1\over 4} H,\qquad Z_2={Q_2\over 4} H,\qquad Q_1=\Omega^2(k^2a^2+k'^2 b^2)Q_5,\\
 \ALM_\phit&=-{Q_5\Omega k \over 2}  \cI_{10},\qquad
 \ALM_\psit =+{Q_5\Omega k'\over 2}  \cI_{01},\\
 \BLM_\phit&= {Q_5\Omega k'\over 2} \left(-\cI_{01}+{1\over 2}b^2H-2\gamma\right),\qquad
 \BLM_\psit = {Q_5\Omega k \over 2} \left(\cI_{10}-{1\over 2}a^2H-2\gamma\right).
\end{split}
\label{jdmf16Apr21}
\end{align}
where $\phit,\psit$ were defined in \eqref{s_w_phit_psit}, and other
components of $\ALM,\BLM$ vanish. The computation is
straightforward, if complicated.  The only thing is that we must smear
over the $\psi$ direction.  For some detail see \eqref{jitw19Apr21}.  The
functions $Z_1,Z_2$ are the same as the one found from the 4d viewpoint
in \eqref{jjyn16Apr21}.  Also, in 6d, the relation \eqref{jsjr14Apr21}
is automatic.  If we set $b\to 0$, \eqref{jdmf16Apr21} reduces to 
\eqref{harm_func_circular}.

We can find what harmonic functions these reduce to, using the results
in section \ref{ss:rel_to_harm_sol}.  We can decompose the 1-forms
$\ALM,\BLM$ as
\begin{subequations} 
 \begin{align}
 \ALM
 &={\ALM_\phit+\ALM_\psit\over 2}(d\psi+A)
 +\left(
 -\ALM_\phit \cos^2\!{\theta\over 2}
 +\ALM_\psit \sin^2\!{\theta\over 2}
 \right)d\phi,
 \\
 \BLM
 &={\BLM_\phit+\BLM_\psit\over 2}(d\psi+A)
 +\left(
 -\BLM_\phit\cos^2\!{\theta\over 2}
 +\BLM_\psit\sin^2\!{\theta\over 2} 
 \right)d\phi.
\label{jdyg16Apr21}
 \end{align}
\end{subequations}
Using the identification \eqref{jtou17Feb21}, we find
\begin{align}
\begin{split}
 M&=-{1\over 4}(\ALM_\phit+\ALM_\psit+\BLM_\phit+\BLM_\psit),\qquad
 K^3={V\over 2}(-\ALM_\phit-\ALM_\psit+\BLM_\phit+\BLM_\psit),\\
 \mu &=-{1\over 2}(\ALM_\phit+\ALM_\psit),\qquad
 \omega=\left(
 \ALM_\phit\cos^2\!{\theta\over 2}
 -\ALM_\psit\sin^2\!{\theta\over 2}\right)d\phi,\\
 \xi^3&=\left[
 (\ALM_\phit-\BLM_\phit)\cos^2{\theta\over 2}
 +(-\ALM_\psit+\BLM_\psit)\sin^2{\theta\over 2}
 \right]d\phi.
\end{split}
\end{align}
It is not difficult to show that these reproduce the harmonic functions
\eqref{K3_final}, \eqref{M_final} in the previous section, using the
formula \eqref{iuds16Apr21}.

The reader may think that, to find harmonic functions, it is much easier
to start with the Lunin-Mathur geometry in 6d and go down to 4d/5d as
above (``top-down''), rather than going from 4d/5d to 6d as we did in
section~\ref{sec:codim-2_LM} (``bottom-up'').  However, a technical
point is that, even if we know $\ALM$, the
relation~\eqref{mfxm17Feb21} only gives us the combination
$\mu=M+K^3/(2V)$, and not $M$ and $V$ separately.  One could use the
harmonicity of $K^3,M$ to disentangle them from each other, but that is
far from simple.  If we also know $\BLM$, it is easy to find
$K^3,M$ from the two relations~\eqref{jtou17Feb21}, but finding $\BLM$ from the duality relation \eqref{def_ALM} is not simple either.
We found the expression for $\BLM$ in
\eqref{jdmf16Apr21} by first finding $K^3,M$ and then going back up to
6d via \eqref{eqnb18Feb21} and \eqref{jdyg16Apr21}.

\section{Dual CFT perspective and spectral flow}
\label{sec:cft}

\subsection{Dual CFT states}

The CFT dual of general Lunin-Mathur geometries are known
\cite{Kanitscheider:2006zf,Kanitscheider:2007wq,Skenderis:2006ah,Giusto:2019qig}. Here
we discuss the dual for the special case of the helical profile,
focusing on its symmetry.

The holographic dual of the D1-D5 system is a 2d SCFT called the D1-D5
CFT, which is an orbifold CFT with target space ${\rm Sym}^N(T^4)$,
$N=N_1 N_5$.\footnote{More precisely the target space is a deformation
of this but we will assume that the target space is this orbifold.}
The CFT has $SU(2)_L\times SU(2)_R$ R-symmetry with generators
$J_L^i,J_R^i$ and the states in the CFT have R-charges
$(J^3_L,J^3_R)=(j_L,j_R)$.  For more detail of the D1-D5 CFT, see
\cite{David:2002wn,Avery:2010qw}. Our notation here follows
\cite{Bena:2017xbt}.

The states of the orbifold CFT can be constructed multiplying ``strands''
of various length together so that the total length is $N$.  The CFT
state that is dual to the circular profile \eqref{profile_circular0} is
\begin{align}
 \bigl[\ket{++}_k\bigr]^{N/k},\qquad N=N_1 N_5.
\label{nfzg17May21}
\end{align}
Here $\ket{++}_k$ is a strand of length $k$, with a Ramond-Ramond ground
state with R-charge $(j_L,j_R)=(\half,\half)$ on it, denoted by
``$++$''.  This state is an eigenstate of $J_L^3,J_R^3$.  The dual bulk
statement is that the supergravity solution preserves the corresponding
$U(1)_L\times U(1)_R$ symmetry.  More precisely, $J_\phit=J_L^3+ J_R^3$ and $J_\psit=J_L^3-J_R^3$
generate rotations in the 1-2 and 3-4 planes under which the
circular profile \eqref{profile_circular} is invariant.

On the other hand, the state dual to the helical profile
\eqref{profile_helix0} is a coherent sum of RR ground states
\cite{Kanitscheider:2006zf,Kanitscheider:2007wq,Skenderis:2006ah,Giusto:2019qig},
\begin{align}
\sideset{}{'} \sum_{N_{++},N_{-+}}
\bigl[A_{++}\,\ket{++}_k\bigr]^{N_{++}}\,
\bigl[A_{-+}\,\ket{-+}_{k'}\bigr]^{N_{-+}}\,,
 \label{hqdc21Apr21}
\end{align}
where the sum is restricted to $(N_{++},N_{-+})$ with
$kN_{++}+k'N_{-+}=N$, and the parameters $A_{++},A_{-+}$ are related to
the bulk quantities $a,b$ via
\begin{align}
 |A_{++}|^2={R_y^2N\over Q_1 Q_5}\,k^2  a^2,\qquad
 |A_{-+}|^2={R_y^2N\over Q_1 Q_5}\,k'^2 b^2.\label{hcxw17May21}
\end{align}
The sum \eqref{hqdc21Apr21} is dominated by the following term:
\begin{align}
 k \overline{N}_{++}=|A_{++}|^2,\qquad
 k' \overline{N}_{-+}=|A_{-+}|^2.\label{gotu17May21}
\end{align}
The state \eqref{hqdc21Apr21} is not an eigenstate of $J_L^3$ and
$J_R^3$ separately, but it is an eigenstate of the combination $
(k-k')J_L^3+(k+k')J_R^3=kJ_\phit-k'J_\psit$.\footnote{One exceptional
case is $k=k'$.  In this case, making $a$ non-vanishing can be thought
of merely as an $SU(2)_L$ rotation of the original state
\eqref{nfzg17May21}. Indeed, one can show that \eqref{nfzg17May21} is an
eigenstate of $J_R^3$ and a certain linear combination of $J_L^i$,
$i=1,2,3$.}  In the bulk, this is nothing but the linear combination of
$\phit$ and $\psit$ directions under which the helical profile is
invariant.  Once we project the profile to $\bbR^3$, this structure
becomes invisible because of smearing; namely, the 4d/5d configuration
is symmetric under both $\phi$ and $\psi$ translations, although this is
an artifact of smearing.

As we keep increasing  $b$, we end up with the $a=0,b>0$ state 
\begin{align}
 \bigl[\ket{-+}_{k'}\bigr]^{N/k'}
\end{align}
which corresponds to the ``north pole'' limit.  This is again
symmetric under $U(1)_L\times U(1)_R$.

\subsection{Spectral flow}

Being an $\cN=(4,4)$ SCFT, the D1-D5 CFT has spectral flow symmetry
\cite{Schwimmer:1986mf} which maps a state with $L_0=h,J_L^3=j_L$ (we
take $L_0=0$ for Ramond ground states) on a strand of length $k$ to a state with
$L_0=h',J_L^3=j_L'$  as
\begin{align}
 h'=h+2 n j_L+k n^2,\qquad 
 j_L'=j_L+kn.
\end{align}
For $n\in\bbZ$, this maps states in the Ramond sector into states in the
Ramond sector.

Let us denote the Ramond state obtained by spectral
flowing $\ket{\pm +}_k$ by $n$ by
\begin{align}
 \ket{\pm +}_{k,n} \qquad\text{with}\qquad
 h=\pm n+kn^2,\quad
 j_L=\pm {1\over 2}+kn.
\end{align}
By spectral flowing the state
\eqref{nfzg17May21}, we obtain 
\begin{align}
 \bigl[\ket{\pm +}_{k,n}\bigr]^{N/k}.\label{iere21May21}
\end{align}
If we spectral flow the state
\eqref{hqdc21Apr21}, we obtain the coherent sum
\begin{align}
\sideset{}{'} \sum_{N_{++},N_{-+}}
\bigl[A_{++}\,\ket{++}_{k,n}\bigr]^{N_{++}}\,
\bigl[A_{-+}\,\ket{-+}_{k',n}\bigr]^{N_{-+}}\,,
\label{hlla17May21}
\end{align}
which has charges
\begin{subequations} 
 \begin{align}
 h &=(n+kn^2)\overline{N}_{++}+(-n+k'n^2)\overline{N}_{-+},
 \label{hkuu17May21}\\
 j_L &=\left({1\over 2}+kn\right)\overline{N}_{++}+\left(-{1\over 2}+k'n\right)\overline{N}_{-+},\\
 j_R &={1\over 2}(\overline{N}_{++}+\overline{N}_{-+}).
 \end{align}
 \label{hcfe17May21}
\end{subequations}

In the bulk, the spectral flow transformation is realized by the
following transformation of  the harmonic functions\cite{Bena:2008wt}
\begin{align}
\begin{gathered}
 \tilde{V}=V+\gamma^3 K^3,\qquad
 \tilde{K}^1=K^1-\gamma^3 L_2,\qquad
 \tilde{K}^2=K^2-\gamma^3 L_1,\qquad
 \tilde{K}^3=K^3,\\
 \tilde{L}_3=L_3-2\gamma^3 M,\qquad 
 \tilde{L}_2=L_2,\qquad
 \tilde{L}_1=L_1,\qquad
 \tilde{M}=M,\qquad
 \tilde{\omega}=\omega,
\end{gathered}
\label{mpwf4May21}
\end{align}
under which the moduli $\tau^I$ defined in \eqref{def_tau^I} transform
as
\begin{align}
 \tau^1\to \tau^1,\qquad
 \tau^2\to \tau^2,\qquad
 \tau^3\to {\tau^3\over {Q_5\Omega\over 2}\gamma^3\tau^3+1}.\label{feqv18May21}
\end{align}
Namely, it is an $SL(2,\bbZ)$ duality transformation of the modulus of
$T^2_{89}$.\footnote{Generalization of this to a three-parameter family
of transformations  with parameters $\gamma^I$,
$I=1,2,3$ gives an $SL(2,\bbZ)$ duality transformation for $\tau^I$ \cite{Bena:2008wt}.  }

Let us study the bulk dual of the spectral flowed states above.  First,
applying \eqref{mpwf4May21} with $\gamma^3=-{2n\over Q_5 \Omega}$ to the codimension-3
harmonic functions \eqref{VKLM_supertube} and doing gauge transformation~\eqref{gauge_transf} with $c^1=-{n\over 2\Omega},c^2=-{n Q_1\over 2 Q_5
\Omega},c^3=0$ so that $K^I$ are $\cO(1/r^2)$ at infinity, we find the
dual of \eqref{iere21May21}:
\begin{align}
  \begin{gathered}
  \tilde V={1+kn\over r}-{kn\over \Sigma},\qquad
  \tilde K^3={Q_5 \Omega k\over 2}\left({1\over \Sigma}-{1\over r}\right),\\
  \tilde K^1={n(kn+1)\over 2\Omega}\left({1\over \Sigma}-{1\over r}\right)
  ,\qquad
  \tilde K^2={Q_1 n(kn+1)\over 2Q_5 \Omega}\left({1\over \Sigma}-{1\over r}\right)
  ,\\
  \tilde L_1={Q_1\over 4}\left({kn+1\over \Sigma}-{kn\over r}\right),\qquad
  \tilde L_2={Q_5\over 4}\left({kn+1\over \Sigma}-{kn\over r}\right),\\
  \tilde L_3=1+{Q_1 n(kn+1)\over 4Q_5 \Omega^2 k }
  \left({kn+1\over \Sigma}-{kn\over r}\right),\qquad
  \tilde M={Q_1\over 16\Omega k}\left({(kn+1)^2\over \Sigma}-{(kn)^2\over r}\right).
 \end{gathered}
\label{jfru21May21}
\end{align}
Next, if we apply the same bulk spectral flow and gauge transformations
to the codimension-2 harmonic functions given by \eqref{jjyn16Apr21},
\eqref{K3_final} and \eqref{M_final}, we obtain
\begin{align}
\begin{gathered}
 \tilde V={1\over r}-{2n\over Q_5 \Omega} K^3,\qquad
 \tilde K^1=-{n\over 2\Omega}{1\over r}+{n\over 2\Omega}H+{n^2\over Q_5 \Omega^2}K^3,\qquad
 \tilde  K^2={Q_1\over Q_5} \tilde K^1,\\
  \tilde K^3=K^3,\qquad
 \tilde L_1={Q_1\over 4}H+{n Q_1\over 2Q_5 \Omega} K^3,\qquad
 \tilde L_2={Q_5\over 4}H+{n \over 2\Omega} K^3,\\
 \tilde L_3=1-{n^2 Q_1\over 4Q_5 \Omega^2}{1\over r}
 +{n^2 Q_1\over 2Q_5 \Omega^2}H+{n^3 Q_1\over 2Q_5^2 \Omega^3}K^3
 +{4n\over Q_5 \Omega}M,\\
 \tilde M=M+{n Q_1\over 8\Omega} H + {n^2 Q_1\over 8Q_5 \Omega^2}K^3
\end{gathered}
\label{mqhg4May21}
\end{align}
where $K^3,M$ are the ones given in \eqref{K3_final}, \eqref{M_final}.
This must be the bulk dual of the CFT state~\eqref{hlla17May21}.
From the $1/r$ fall-off of $\tilde{L}_3$ and $\tilde{M}$, we can confirm
the charges \eqref{hcfe17May21}.  For example, the large $r$
expansion of $\tilde{L}_3$ is
\begin{align}
 \tilde{L}_3  
 &\sim \left[-{n^2Q_1\over 4Q_5\Omega^2}
 +{n^2Q_1\over 2Q_5\Omega^2}+{4n\over Q_5\Omega}{Q_5\Omega\over 4}
 {1\over 4}(ka^2-k'b^2)\right]{1\over r}
\notag\\
&
 =
{1\over 4r}\Bigl[(n+kn^2)ka^2+(-n+k'n^2)k'b^2\Bigr]
 \equiv {Q_p\over 4r},
\end{align}
where we used \eqref{jsjr14Apr21}.  Using \eqref{hcxw17May21} and
\eqref{gotu17May21}, and using the relation
between $Q_p$ and the
quantized momentum $N_p$,
\begin{align}
 Q_p
 ={g_s^2 \ap^4 \over R_y^2 v_4}N_p
 ={Q_1Q_5\over R_y^2N}N_p,
\end{align}
it is easy to show that this reproduces the CFT
result~\eqref{hkuu17May21}.

\begin{figure}
 \begin{center}
 \begin{tikzpicture}[scale=2]
  \pgfmathsetmacro{\ys}{0.2} 

  \draw[-latex]  (0,0) -- (0,8*\ys) node [above] {$L_0$};
  \draw[-latex]  (-3,0) -- (3,0) node [right] {$J_L^3$};

  \clip (-3,-.5) rectangle (3,8*\ys); 

  \draw[fill=black!5!white]
  (-7/2,12*\ys) -- (-5/2,6*\ys) -- (-3/2,2*\ys) -- (-1/2,0)
  -- (1/2,0) -- (3/2,2*\ys) -- (5/2,6*\ys) -- (7/2,12*\ys);

  \draw[-latex]  (0,0) -- (0,8*\ys) node [above] {$L_0$};
  \draw[-latex]  (-3,0) -- (3,0) node [right] {$j_L\over N$};

  \begin{scope}[very thick,decoration={
    markings,
    mark=at position 0.6 with {\arrow{latex}}}
    ] 
  \draw[postaction={decorate}] (1/2,0) -- (-1/2,0);
  \draw[postaction={decorate}] (-1/2,0) -- (-3/2,2*\ys);
  \draw[postaction={decorate}] (-3/2,2*\ys) -- (-5/2,6*\ys);
  \draw[postaction={decorate}] (-5/2,6*\ys) -- (-7/2,12*\ys);
  \end{scope}

  \draw[dashed] (-5/2,0) -- +(0,6*\ys);
  \draw[dashed] (-3/2,0) -- +(0,2*\ys);
  \draw[dashed] (5/2,0) -- +(0,6*\ys);
  \draw[dashed] (3/2,0) -- +(0,2*\ys);

  \node () at (0,0) [below] {$0$};
  \node () at (1/2,0) [below] {$N\over 2$};
  \node () at (-1/2,0) [below] {$-{N\over 2}$};
  \node () at (3/2,0) [below] {$3N\over 2$};
  \node () at (-3/2,0) [below] {$-{3N\over 2}$};
  \node () at (5/2,0) [below] {$5N\over 2$};
  \node () at (-5/2,0) [below] {$-{5N\over 2}$};

 \fill[black] (1/2,0) circle (0.04);
 \fill[black] (-1/2,0) circle (0.04);
 \fill[black] (-3/2,2*\ys) circle (0.04);
 \fill[black] (-5/2,6*\ys) circle (0.04);
   
  \node () at (1/2,0) [above,xshift=-5,yshift=0] {\footnotesize $\ket{++}_1$};
  \node () at (-1/2,0) [above,xshift=7,yshift=0] {\footnotesize $\ket{++}_{1,-1}$};
  \node () at (-3/2,2*\ys) [above right,xshift=-2,yshift=-3] {\footnotesize $\ket{++}_{1,-2}$};
  \node () at (-5/2,6*\ys) [above right,xshift=-2,yshift=-3] {\footnotesize $\ket{++}_{1,-3}$};
 \end{tikzpicture}
  \vspace*{-1cm}
 \end{center}
 \caption{Repeated interpolations in CFT to reach $[\ket{++}_{1,-m}]^N$ with arbitrary $m$.\label{fig:repeat_flows}
 }
\end{figure}
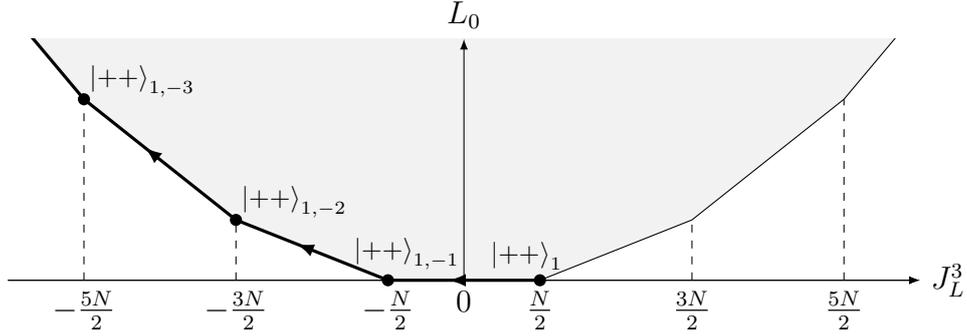

So, the harmonic solution \eqref{mqhg4May21} provides the supergravity
description of the CFT state~\eqref{hlla17May21}. What is interesting is
that, because $\tilde{V}$ contains the monodromic $K^3$ that can
transfer charge, the solution interpolates between Gibbons-Hawking
spaces with different Taub-NUT charges.
\begin{align}
 \tilde{V}={1+kn\over r}-{kn\over \Sigma}
 ~~~~\to~~~~
 \tilde{V}={1-k'n\over r}+{k'n\over \Sigma'},
\end{align}
where $\Sigma'=|\yv-{\bf \bt}|$, ${\bf \bt}=(0,0,\bt)$, $\bt=b^2/4$.  In
particular, if $k=k'=1$, $n=-m$, this connects the two configurations:
\begin{align}
 \tilde{V}={1-m\over r}+{m\over \Sigma}
 ~~~~\to~~~~
 \tilde{V}={1+m\over r}-{m\over \Sigma'}
 ={1-(m+1)\over \Sigma'}+{m+1\over r}.
\label{iafd21May21}
\end{align}
On the right hand side, by reinterpreting $\Sigma'\to r,r\to \Sigma$, we
can repeat the same process now with $m\to m+1$.  Doing this over again, we
can reach any $m$.  See Figure~\ref{fig:repeat_bulk} for a graphical
explanation of this process.
In CFT, the process
\eqref{iafd21May21}
corresponds to the interpolation
\begin{align}
 [\ket{++}_{1,-m} ]^N
 ~~~~\to~~~~
 [\ket{-+}_{1,-m}]^N
 =[\ket{++}_{1,-(m+1)}]^N,
\end{align}
where in the last expression we used the fact that
$\ket{-+}_{1,-m}=[\ket{++}_{1,-(m+1)}]$.  So, by starting from
$[\ket{++}_1]^N=[\ket{++}_{1,0}]^N$ and repeatedly doing this
interpolation, we can get to $ [\ket{++}_{1,-m} ]^N$ with any $m$; see
Figure~\ref{fig:repeat_flows} for a description of this spectral flow on
the $J_L^3$-$L_0$ plane.

\begin{figure}
  \begin{center}
 \begin{tikzpicture}[scale=1]
  \draw[dashed] (0,0) circle (1);
  \fill[green!50!black] (0,0) circle (0.05) node [above] {\footnotesize $1-m$};
  \fill[magenta] (0,-1) circle (0.05) node [below] {\footnotesize $m$};

  \tikzset{shift={(2,0)}};
  \draw[-latex,ultra thick] (-0.3,0) -- (0.3,0);
  \tikzset{shift={(2,0)}};

  \draw[dashed] (0,0) circle (1);
  \fill[green!50!black] (0,0) circle (0.05) node [above] {\footnotesize $1+m$};
  \fill[magenta] (0,1) circle (0.05) node [above] {\footnotesize $-m$};

  \tikzset{shift={(2,0.5)}};
  \node () at (0,0) {$=$};
  \tikzset{shift={(2,0.5)}};

  \draw[dashed] (0,0) circle (1);
  \fill[green!50!black] (0,0) circle (0.05) node [above] {\footnotesize $1-(m+1)$};
  \fill[magenta] (0,-1) circle (0.05) node [below] {\footnotesize $m+1$};

  \tikzset{shift={(2,0)}};
  \draw[-latex,ultra thick] (-0.3,0) -- (0.3,0);
  \tikzset{shift={(2,0)}};

  \draw[dashed] (0,0) circle (1);
  \fill[green!50!black] (0,0) circle (0.05) node [above] {\footnotesize $1+(m+1)$};
  \fill[magenta] (0,1) circle (0.05) node [above] {\footnotesize $-(m+1)$};
 \end{tikzpicture}
  \vspace*{-.7cm}
 \end{center}
 \caption{\label{fig:repeat_bulk} Repeated interpolations in the bulk.}
\end{figure}
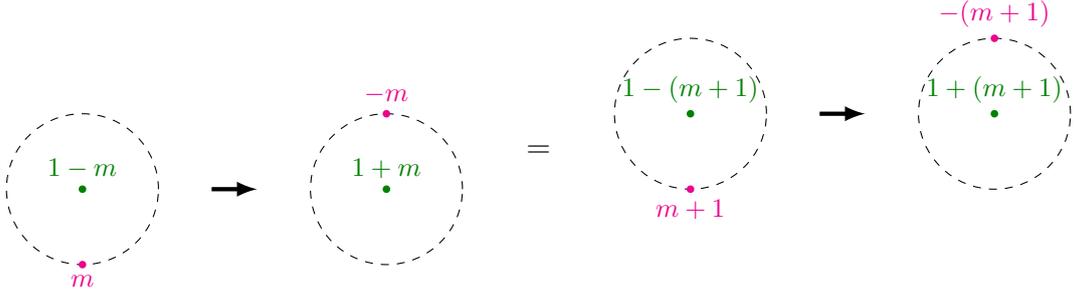

From \eqref{fenv18May21} and \eqref{feqv18May21}, we see that the
modulus $\tau^3$ now has the $SL(2,\bbZ)$ monodromy
\begin{align}
\tilde\cM= \begin{pmatrix}
	      1-n(k+k') & -(k+k')\\
	     n^2(k+k') & 1+n(k+k')
	    \end{pmatrix}.
\end{align}
This is no longer the monodromy of NS5(4567$\cC$)-branes but the
``exotic brane'' $5^2_2(4567\cC,89)$ \cite{deBoer:2012ma} has mixed
in. By following the duality transformation to the D1-D5 frame, we see
that this is mapped as
\begin{align}
 5^2_2(4567\cC,89)
 \xrightarrow{~\text{dualize}~}
 {\rm KKM}(6789\cC,\psi).
\end{align}
So, in addition to the KKMs in \eqref{D1+D5->KKM2}, we have new a kind
of KKM dipole, but the solution is purely geometric.  This is as it
should be, because the spectral flow transformation in the D1-D5 frame
is nothing but mixing the coordinates $y$ and $\psi$; part of the KKM in
the second line of \eqref{D1+D5->KKM2} got transformed into ${\rm
KKM}(6789\cC,\psi)$ above.

\subsection{Fractional spectral flow}
For general CFT states, the spectral flow is defined only for $n\in\bbZ$.
However, for the state $\ket{\pm +}_k$, the fractional spectral flow by $n={s\over k}$, $s\in\bbZ$,
\begin{align}
 \ket{\pm +}_{k,{s\over k}}:
 \qquad h={s(s\pm 1)\over k},\quad
 j_L=\pm {1\over 2 }+s
\end{align}
is also a valid state \cite{Giusto:2012yz} if
\begin{align}
h= {s(s\pm 1)\over k}\in\bbZ,\qquad
\text{or}\qquad {n(nk\pm 1)}
 \in\bbZ.\label{jfvr21May21}
\end{align}
Indeed, the dual bulk solution \eqref{jfru21May21} is known to make
sense \cite{Giusto:2012yz} if this condition \eqref{jfvr21May21} is met,
and represents the most general 2-center codimension-3 harmonic
solution.

The interpolating CFT state \eqref{hlla17May21} must also be a valid
state if both the states $\ket{++}_{k,n}$ and $\ket{-+}_{k',n}$ satisfy
the quantization condition, namely,
\begin{align}
 n(nk+1)\in\bbZ\qquad\text{and}\qquad  n(nk'-1)\in\bbZ.
\end{align}
For example, for $m,m'\in\bbZ_{>0}$,  
the following states with $n=1/2$:
\begin{align}
&\ket{++}_{4m+2,{1\over 2}},
 \quad
 h=m+1\in\bbZ,
 \quad j_L=2m+{3\over 2},\\
&\ket{-+}_{4m'-2,{1\over 2}},
 \quad
 h=m'-1\in\bbZ,
 \quad j_L=2m'-{3\over 2}
\end{align}
satisfy the quantization condition and the interpolation will be an
allowed state.  The dual geometry is still given by \eqref{mqhg4May21}.

\section{Discussions}
\label{sec:discussions}

In this note, we studied the Lunin-Mathur geometry interpolating the
states
\begin{align}
 [\ket{++}_{k}]^{N/k}
~~~~\text{and}~~~~
 [\ket{-+}_{k'}]^{N/k'}
\label{eutc22May21}
\end{align}
in the framework of harmonic solutions in 4d/5d.  Although the
geometries dual to \eqref{eutc22May21} are codimension-3 solutions, the
interpolating solution is of codimension two, because of the puffed-out
NS5-branes lying along a curve $\cC$.  We discussed how to construct the
associated harmonic functions with a monodromy around $\cC$, based on
the data about the charges and the puffed-out dipole charge.
The interpolating solution exhibits some interesting features, such as
some of the D4-charge being delocalized, and some of the D4-charge
getting transferred from the supertube center to the Taub-NUT center as
the interpolation proceeds.
We also discussed the spectral flow of this entire process, which
interpolates between the states
\begin{align}
 [\ket{++}_{k,n}]^{N/k}
~~~~\text{and}~~~~
 [\ket{-+}_{k',n}]^{N/k'}.
\end{align}
This solution is valid even if $n$ is not an integer, as long as
certain quantization conditions are met; in that case, the interpolating
solution connects the 2-center solutions found in 
\cite{Giusto:2012yz}.

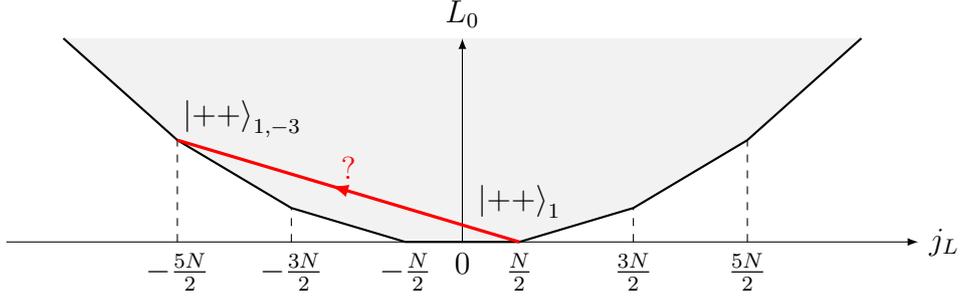
\begin{figure}
 \begin{center}
 \begin{tikzpicture}[scale=1.5]
  \pgfmathsetmacro{\ys}{0.15} 



\draw[thick,fill=black!5!white] (-7/2,12*\ys) -- (-5/2,6*\ys) -- (-3/2,2*\ys) -- (-1/2,0) -- (1/2,0) -- (3/2,2*\ys) -- (5/2,6*\ys) -- (7/2,12*\ys);

  \draw[-latex]  (0,0) -- (0,12*\ys) node [above] {$L_0$};
  \draw[-latex]  (-4,0) -- (4,0) node [right] {$j_L$};

  \draw[dashed] (-5/2,0) -- +(0,6*\ys);
  \draw[dashed] (-3/2,0) -- +(0,2*\ys);
  \draw[dashed] (5/2,0) -- +(0,6*\ys);
  \draw[dashed] (3/2,0) -- +(0,2*\ys);

  \node () at (0,0) [below] {$0$};
  \node () at (1/2,0) [below] {$N\over 2$};
  \node () at (-1/2,0) [below] {$-{N\over 2}$};
  \node () at (3/2,0) [below] {$3N\over 2$};
  \node () at (-3/2,0) [below] {$-{3N\over 2}$};
  \node () at (5/2,0) [below] {$5N\over 2$};
  \node () at (-5/2,0) [below] {$-{5N\over 2}$};
  
  \begin{scope}[very thick,red,decoration={
    markings,
    mark=at position 0.55 with {\arrow{latex}}}
    ] 
  \draw[postaction={decorate}] (1/2,0) -- (-5/2,6*\ys) node [midway,above] {?};
  \end{scope}
  
  \node () at (1/2,0) [above,yshift=5] {$\ket{++}_1$};
  \node () at (-5/2,6*\ys) [above right,xshift=-2,yshift=-3] {$\ket{++}_{1,-3}$};
  
 \end{tikzpicture}
 \end{center}
 \caption{\label{fig:puzzle} An example of interpolation between
 $[\ket{++}_{k,n}]^{N/k}$ and $[\ket{++}_{k',n'}]^{N/k'}$ with $n\neq n'$.}
\end{figure}

A natural question 
\cite{Hampton:2018ygz}
is how to interpolate between the
states
\begin{align}
 [\ket{++}_{k,n}]^{N/k}
~~~~\text{and}~~~~
 [\ket{-+}_{k',n'}]^{N/k'}
\end{align}
where $n\neq n'$.  For definiteness, let us set $k=k'=1,n=0$, and $n'\to
n+1$.  Then the interpolation in question is between
\begin{align}
 [\ket{++}_{1}]^{N}
~~~~\text{and}~~~~
 [\ket{-+}_{1,n+1}]^{N}
 =[\ket{++}_{1,n}]^{N}.
\end{align}
See Figure \ref{fig:puzzle} for an example.  In CFT, it is easy to write down the
interpolating state, as in~\eqref{hlla17May21}:
\begin{align}
\sideset{}{'} \sum_{N_{++},N_{-+}}
\bigl[A_1\,\ket{++}_{1}\bigr]^{N_{++}}\,
\bigl[A_n\,\ket{++}_{1,n}\bigr]^{N_{-+}}\,.
\label{epwf21Jun21}
\end{align}
Just as \eqref{hlla17May21}, this is an eigenstate of $J_R$ and a
certain linear combination of $J_L$ and $J_R$.  This suggests that the
bulk dual has a helical structure similar to the helical Lunin-Mathur
profile that we studied.  When reduced to $\bbR^3$, it is likely to be
described by a solution involving codimension-2 sources similar to the
one studied in this note.  Constructing such solutions would be very
interesting for understanding black hole microstates, because the state
$\ket{++}_{k,n}$ in the background of $\ket{++}_{1}$ is a stringy state
in AdS$_3\times S^3$ \cite{Lunin:2002fw,Gomis:2002qi,Gava:2002xb}, which
is a key for understanding the fractional and higher modes responsible
for the entropy.  The technique developed in this note should be useful
for such construction. Note, however, that the interpolating states in
CFT such as \eqref{epwf21Jun21} are non-BPS when perturbed away from the
orbifold point \cite{Gava:2002xb,Hampton:2018ygz}.  Our harmonic
function techniques will be directly useful only if such states are
actually BPS in supergravity.  Such a possibility is not outrageous,
because of the observed correspondence between BPS states in the CFT at
the orbifold point and multi-center solutions in supergravity
\cite{Bena:2011zw}.

Relatedly, in \cite{Heidmann:2018vky}, the moduli space of the multi-center
codimension-3 solutions for small numbers of D6- and D2-branes was
investigated.  They found that, for one D6- and three D2-branes, the
number of solutions agrees with the prediction from quiver quantum
mechanics, although for more D6-branes the number of multi-center
solutions is less than the prediction.  It would be interesting to
explore the moduli space of multi-center solutions including
codimension-2 sources.  The results in this note should be useful 
for such exploration.

\section*{Acknowledgments}

I would like to thank CEA Saclay for their (online) hospitality in the
``Black-Hole Microstructure I \& II'' workshops.  This work was
supported in part by MEXT KAKENHI Grant Numbers 17H06357 and 17H06359.

\appendix

\section{Duality transformation}
\label{app:duality}

Here we discuss how to map the IIB configuration in \eqref{D1+D5->KKM2}
in the D1-D5 frame into the IIA configuration \eqref{D2+D2->D4+NS5}.

We start with \eqref{D1+D5->KKM2}:
\begin{align}
 \text{D1}(y) + \text{D5}(y6789)
 \xrightarrow{\text{puff out}} 
\begin{cases}
 \text{KKM}(6789\psi,y)+\text{P}(\psi)\\[.5ex]
 \text{KKM}(6789\cC,y)+\text{P}(\cC)
\end{cases}
\end{align}
After T-duality along $y67$, this is mapped into
\begin{align}
 \text{D2}(67) + \text{D2}(89)
 \xrightarrow{\text{puff out}} 
\begin{cases}
 \text{NS5}(6789\psi)+\text{P}(\psi)\\[.5ex]
 \text{NS5}(6789\cC)+\text{P}(\cC)
\end{cases}
\end{align}
Lifting this to M-theory along $x^{10}$, this becomes
\begin{align}
 \text{M2}(67) + \text{M2}(89)
 \xrightarrow{\text{puff out}} 
\begin{cases}
 \text{M5}(6789\psi)+\text{P}(\psi)\\[.5ex]
 \text{M5}(6789\cC)+\text{P}(\cC)
\end{cases}
\end{align}
Renaming coordinates as $(6,7,8,9,y,10)\to (4,5,6,7,8,9)$, we have
\begin{align}
 \text{M2}(45) + \text{M2}(67)
 \xrightarrow{\text{puff out}} 
\begin{cases}
 \text{M5}(4567\psi)+\text{P}(\psi)\\[.5ex]
 \text{M5}(4567\cC)+\text{P}(\cC)
\end{cases}
\end{align}
This is a supertube transition in the M-theory frame of \eqref{Msol}.
Compactifying M-theory to IIA on $\psi$, this reduces to
\begin{align}
 \text{D2}(45) + \text{D2}(67)
 \xrightarrow{\text{puff out}} 
\begin{cases}
 \text{D4}(4567)+\text{D0}\\[.5ex]
 \text{NS5}(4567\cC)+\text{P}(\cC)
\end{cases}
\end{align}
This is the process \eqref{D2+D2->D4+NS5} in the IIA frame of \eqref{IIAfield}.

\section{Coordinate systems}

Here we summarize relations between the coordinate systems that we use
in the main text.

The flat $\bbR^4$ coordinates $x_m$, $m=1,2,3,4$, are variously written as
\begin{align}
\begin{array}{l@{~}l@{~}l@{~}l@{~}l}
 x_1+ix_2&=&se^{i\phit}&=&\displaystyle 2\sqrt{r}\,\sin\tfrac{\theta}{2}\,e^{i{\psi\over 2}},\\[1ex]
 x_3+ix_4&=&we^{i\psit}&=&\displaystyle 2\sqrt{r}\,\cos\tfrac{\theta}{2}\,e^{i({\psi\over 2}+\phi)}.
\end{array}
\end{align}
The flat $\bbR^4$ metric can be written as
\begin{align}
\begin{split}
  ds_4^2
 &=dx_m dx_m\\
 &=ds^2+ s^2 d\phit^2+dw^2+ w^2 d\psit^2\\
 &={1\over r}dr^2+r\left(d\theta^2+2(1-\cos\theta)d\phit^2+2(1+\cos\theta)d\psit^2\right)\\
 &=V^{-1}(d\psi+\cA)^2+V d\yv^2,\\
 \cA&=(1+\cos\theta)d\phi,\quad V={1\over r},\quad *_3 d\cA=dV,\\
\end{split}
\end{align}
where
\begin{align}
 d\yv^2&=dr^2+r^2(d\theta^2+\sin^2\!\theta\,d\phi^2),\\
 y_1+iy_2&=r\sin\theta\, e^{i\phi},\qquad
 y_3=r\cos\theta.
\end{align}

In the toroidal coordinates
\begin{align}
 y_1+iy_2={\sqrt{u^2-1}\over u-\cos\sigma}Re^{i\phi},\qquad
 y_3'\equiv y_3-c={\sin\sigma\over u-\cos\sigma}R,
\end{align}
the flat $\bbR^3$ metric can be written as
\begin{align}
 d\yv^2&=
 d\yv'^2=
dy_1^2+dy_2^2+dy_3'{}^2={R^2\over (u-\cos\sigma)^2}
 \left[{du^2\over u^2-1}+{d\sigma^2}+(u^2-1)d\phi^2\right]
\end{align}
where $\yv'=(y_1,y_2,y_3')$.
The inverse relations are 
\begin{align}
 \cos\sigma={\yv'^2-R^2\over \Lambda},\qquad
 u={\yv'^2+R^2\over \Lambda},\qquad
 \Lambda^2=(\yv'^2-R^2)^2+4R^2 y_3'{}^2.
\end{align}
Some relations between different coordinates:
\begin{subequations} 
 \begin{gather}
 r={s^2+w^2\over 4}={1\over 4}|\vec{x}|^2,\qquad
 \sin\frac{\theta}{2}={s\over \sqrt{s^2+w^2}},\qquad
 \cos\frac{\theta}{2}={w\over \sqrt{s^2+w^2}},\\
 \cos{\theta}={-s^2+w^2\over s^2+w^2},\qquad
 \sin{\theta}={2sw\over s^2+w^2}\\
  s=2\sqrt{r}\sin{\theta\over 2}=\sqrt{2r(1-\cos\theta)}=\sqrt{2(r-y_3)}
 ,\\ w=2\sqrt{r}\cos{\theta\over 2}=\sqrt{2r(1+\cos\theta)}=\sqrt{2(r+y_3)}
  \\
 \rho=\sqrt{y_1^2+y_2^2}=r\sin\theta={sw\over 2},\qquad
 y_3=r\cos\theta ={-s^2+w^2\over 4}.\label{fsfz13Apr21}
 \end{gather}
\end{subequations}
The relation between the position of the ring in $\bbR^4$ (specified by $a,b$) and in
$\bbR^3$ (specified by $R,c$):
\begin{align}
 R={ab\over 2},\quad c={-a^2+b^2\over 4},
 \qquad
 \sqrt{R^2+c^2}={a^2+b^2\over 4}.
 \label{fsgo13Apr21}
\end{align}
Some more relations: 
\begin{subequations} 
\begin{align}
 V&={1\over r}={4\over s^2+w^2}={4\over |\vec{x}|^2}\label{fsgx13Apr21}
 \\
 \Lambda^2&=
 {1\over 256}
 [(s+a)^2+(w+b)^2]
 [(s+a)^2+(w-b)^2]
 \notag\\[-1ex]
 &\hspace{10ex}
 \times [(s-a)^2+(w+b)^2]
 [(s-a)^2+(w-b)^2]\\[1ex]
 u&=\tfrac{(s^2+w^2)^2+(a^2+b^2)^2-2(a^2-b^2)(s^2-w^2)}{\sqrt{[(s+a)^2+(w+b)^2][(s+a)^2+(w-b)^2][(s-a)^2+(w+b)^2][(s-a)^2+(w-b)^2]}},\\[1ex]
 \zeta&=\sqrt{1-u^{-2}}=\tfrac{8absw}{(s^2+w^2)^2+(a^2+b^2)^2-2(a^2-b^2)(s^2-w^2)},
  \\[1ex]
 \Sigma&={1\over 4}\sqrt{[(s+a)^2+w^2][(s-a)^2+w^2]}
 =|\yv-{\bf \at}|,\\
 {\bf \at}&=(0,0,-\at),\qquad
 \at={a^2\over 4},\qquad \bt={b^2\over 4}
\end{align}
\end{subequations}

\section{Functions $H$, $\cI_{mn}$, $\gamma$}
\label{app:funcs}

Here we summarize the functions introduced in the main text and discuss
their properties.

\subsection{The harmonic function $H$}

The harmonic function $H$ can be written in various ways as
\begin{align}
\begin{split}
  H&={1\over R}\sqrt{u-\cos\sigma\over 2u}\,\,_{2}F_{1}\!\left({1\over 4},{3\over 4};1;\zeta^2\right)
 ={2\over \pi R}\sqrt{{u-\cos\sigma\over 2u}}\,{{\bf K}({2\zeta\over \zeta-1})\over \sqrt{1-\zeta}}\\
 &={1\over R}\sqrt{{u-\cos\sigma\over 2}}\, P_{-1/2}(u)
=
{4\,\cI_{00}(P,Q)\over s^2+a^2+w^2+b^2}.
\label{cec20Apr21}
\end{split}
\end{align}
where $\zeta=\sqrt{1-u^{-2}}$ and ${\bf K}(m)$ is the complete elliptic
integral of the first kind.  The value of $H$ at $r=0$ is easy to see
from the last expression, because $r=0$ means $P=Q=0$:
\begin{align}
 H(r=0)={4\over a^2+b^2}
\end{align}
The behavior of $H$ near the ring and near infinity is given in 
\eqref{nfha8Apr21}.

$H$ appears in the functions $Z_{1,2}$ of the Lunin-Mathur geometry for
the helical profile \eqref{profile_helix} after smearing.  For example, $Z_2$ is
\begin{align}
 Z_2
 &={Q_5\over L}\int_0^{L}{d\lambda\over |\vec x-\vec g|^2}
\notag\\
 &={Q_5\over L}\int_0^{2\pi}
 {d\lambda\over s^2+a^2+w^2+b^2-2sa\cos(\phit-k\Omega \lambda) -2wb\cos(\psit+k'\Omega \lambda) }.
\end{align}
After smearing along $\psi$, this becomes
\begin{align}
\begin{split}
  Z_2
 &=
 Q_5 \int_0^{2\pi} {d\mu\over 2\pi}
 \int_0^{2\pi} {d\mu\over 2\pi}
 {1\over s^2+a^2+w^2+b^2-2sa\cos\mu -2wb\cos\nu }
 \\
 &=
 {Q_5\over s^2+a^2+w^2+b^2}\int_0^{2\pi}{d\mu\over 2\pi}
 \int_0^{2\pi}{d\nu\over 2\pi}
 {1\over 1-P\cos\mu - Q\cos\nu}
 \\
 &
 =
 {Q_5\over s^2+a^2+w^2+b^2}\cI_{00}(P,Q)
 ={Q_5\over 4}H.
\end{split}
\label{jitw19Apr21}
\end{align}
We can relate this to an integral over a 3d profile.  Let us set
$\nu\to\nu+\mu$
in the second expression of \eqref{jitw19Apr21} so that
\begin{equation}
\begin{gathered}
  P\cos\mu+Q\cos\nu
 \to 
 P\cos\mu+Q\cos(\nu+\mu)
 =S\cos(\mu+\mu_0),\\
 S^2=P^2+Q^2+2PQ\cos\nu,\qquad
 \cos\mu_0={P+Q\cos\nu\over S},\qquad
 \sin\mu_0={Q\sin\nu\over S}.
\end{gathered}
\end{equation}
If we use this and carry out the $\mu$ integral, we find
\begin{equation}
\begin{split}
  Z_2
 &={Q_5\over s^2+a^2+w^2+b^2}
 \int_0^{2\pi}{d\nu\over 2\pi}
 {1\over \sqrt{1-S^2}}\\
 &={Q_5\over s^2+a^2+w^2+b^2}
 \int_0^{2\pi}{d\nu\over 2\pi}
 {1\over \sqrt{1-(P^2+Q^2)-2PQ\cos\nu}}.
\end{split}
\end{equation}
We can show that this is equal to  the integral over the 3d profile:
\begin{align}
 Q_5\int_0^{2\pi} {d\nu\over 2\pi}{1\over |\yv-\fv|}
 ={Q_5\over 4}
 \int {d\nu\over 2\pi}
 {1\over \sqrt{\rho^2+R^2-2\rho R\cos\nu + (y_3-c)^2}}
\end{align}
where $\fv$ is given by \eqref{njrp11Apr21}, because, from
\eqref{fsfz13Apr21} and \eqref{fsgo13Apr21}, it follows that
\begin{align}
 \rho^2+R^2-2\rho R\cos\nu + (y_3-c)^2
 &={(s^2+w^2+a^2+b^2)^2\over 16}[1-(P^2+Q^2)-2PQ\cos\nu].
\end{align}
This means that $H$ can be written as an integral over a 4d or 3d
profile as
\begin{equation}
H=
{4\over L}\left.\int_0^L {d\lambda\over |\vec{x}-\vec{g}|^2}\right|_{\rm smear}
=\int_0^{2\pi} {d\nu\over 2\pi}{1\over |\yv-\fv|}.
\end{equation}

%
%

\subsection{The integrals $\cI_{mn}$}

$\cI_{mn}$ is defined as
\begin{subequations}
 \begin{gather}
 \cI_{mn}(P,Q)\equiv 
 \int_0^{2\pi}{d\mu\over 2\pi}
 \int_0^{2\pi}{d\nu\over 2\pi}\,
 {(P\cos\mu)^m (Q\cos\nu)^n\over 1-P\cos\mu - Q\cos\nu},
\label{def_Imn_app}
\\
 P\equiv {2sa\over s^2+a^2+w^2+b^2},\qquad
 Q\equiv {2wb\over s^2+a^2+w^2+b^2}.
\label{def_PQ_app}
 \end{gather}
\end{subequations}
The value at $r=0$ is easy to find because $r=0$ means $P=Q=0$.  Namely,
\begin{align}
 \cI_{mn}(r=0)=
\begin{cases}
 1 &\quad (m=n=0),\\
 0 &\quad \text{(otherwise).}
\end{cases}
\end{align}

One can evaluate $\cI_{00}$ as follows.  First,
\begin{align}
 \cI_{00}
 &=\int_0^{2\pi}{d\mu\over 2\pi}
 \int_0^{2\pi}{d\nu\over 2\pi}
 {1\over 1-P\cos\mu -Q\cos\nu}
 =\int_0^{2\pi}{d\nu\over 2\pi}
 {1\over \sqrt{(1-Q\cos\nu)^2 -P^2}}
 \notag\\
 &=
\oint {dx\over 2\pi i}
 {1\over \sqrt{(x-(Q/2)(x^2+1))^2-P^2x^2}},\qquad
 x=e^{i\nu},
\end{align}
where the contour for the $x$ integral is the unit circle on the complex
$x$ plane.  The roots of the quartic polynomial inside the square root
are $x=x_a,x_b,x_c,x_d$ with
\begin{gather}
 x_d=\tfrac{1+P-\sqrt{(1+P)^2-Q^2}}{4Q},\quad
 x_c=\tfrac{1-P-\sqrt{(1-P)^2-Q^2}}{4Q},\\
 x_b=\tfrac{1-P+\sqrt{(1-P)^2-Q^2}}{4Q},\quad
 x_a=\tfrac{1+P+\sqrt{(1+P)^2-Q^2}}{4Q},
\end{gather}
where $0<x_d<x_c<1<x_b<x_a$ for $P,Q,P+Q\in(0,1)$.
From this it immediately follows that \cite[3.147.2]{GR}
\begin{align}
\cI_{00}
&={4\over \pi}{{\bf K}(m)\over  Q\sqrt{(x_a-x_c)(x_b-x_d)}},
\end{align}
where
\begin{gather}
 m={(x_a-x_b)(x_c-x_d)\over (x_a-x_c)(x_b-x_d)}
 =\left(
 \tfrac{\sqrt{1-(P-Q)^2}-\sqrt{1-(P+Q)^2}}
 {\sqrt{1-(P-Q)^2}+\sqrt{1-(P+Q)^2}}
\right)^2,\\
 {1\over Q\sqrt{(x_a-x_c)(x_b-x_d)}}
 ={1\over \sqrt{1-(P+Q)^2}+\sqrt{1-(P-Q)^2}}.
\end{gather}
By using the identity 
\cite{functions.wolfram}
\begin{align}
 {\bf K}(z)=\tfrac{2}{1+\sqrt{1-z}}\,{\bf K}\Bigl(\Bigl(\tfrac{1-\sqrt{1-z}}{1+\sqrt{1-z}}\Bigr)^2\Bigr)
\end{align}
one can obtain the second expression in
\eqref{cec20Apr21}.

Likewise, we can write $\cI_{01},\cI_{10}$ using elliptic integrals.  For example,
\begin{align}
\cI_{01}
 &={1\over \pi}
 \int_{x_d}^{x_c}  dx
 {x\over \sqrt{(x_a-x)(x_b-x)(x_c-x)(x-x_d)}}
\notag\\
 &={4\over \pi}
 {1\over \sqrt{(x_a-x_c)(x_b-x_d)}}
\left[
 (x_d-x_a)\,\Pi\bigl(\tfrac{x_d-x_c}{x_a-x_c}\big|m\bigr)
 +x_a{\bf K}(m)
 \right]
\end{align}
where $\Pi(n|m)$ is the complete elliptic integral of the third kind.

The 1-form $\ALM$ in the Lunin-Mathur geometry can be written in the $(s,\phit,w,\psit)$ coordinates in
terms of $\cI_{10},\cI_{01}$ as
\begin{align}
\begin{split}
 A_\phit&=\Im[(A_1+iA_2)e^{-i\phit}]s
 =-{Q_5 \Omega k\over 2} \cI_{10},\\
 A_\psit&=\Im[(A_3+iA_4)e^{-i\psit}]w
 =+{Q_5\Omega k'\over 2} \cI_{01}.
\end{split}
\end{align}

\subsection{The monodromic harmonic function $\gamma$}

The explicit expression for $\gamma$ was given in \eqref{egzm3Mar21}.

The value of $\gamma$ on the $y_3$ axis, namely $u=1$, can be found by
using the explicit expression~\eqref{egzm3Mar21} as
\newcommand{\floor}[1]{{\left\lfloor{#1} \right\rfloor}}
 \begin{align}
 \gamma(u=1,\sigma)
 =\sin^2\!\left({\pi\over 2}\left\{{\sigma\over 2\pi}\right\}\right)+\floor{\sigma\over 2\pi}
\label{gamma(u=1)}
 \end{align}
where $\floor{x}$ is the floor function and $\{x\}\equiv x-\floor{x}$ is
the fractional part of $x$.

Let us find the value of $\gamma$ at $r=0$. On the $y_3$ axis, the
relation between $\sigma$ and $y_3'=y-c$ is $y_3'=R\cot(\sigma/2)$, as
can be derived from \eqref{hdtf19Mar21}.  Using this, we can show that
the value of $\gamma$ at $r=0$ ($y_3'=-c=(a^2-b^2)/4$) is
\begin{align}
 \gamma(r=0)\equiv {b^2\over a^2+b^2}+n,
\end{align}
where $n\in\bbZ$ must be taken appropriately for the branch in
consideration.  Let us consider the branch
\begin{align}
 -\pi < \sigma < \pi,\qquad\text{therefore}\quad
 -{1\over 2}<\gamma(u=1,\sigma)<{1\over 2}.
\end{align}
With this choice, $\gamma$ vanishes at infinity (which corresponds to
$u=1,\sigma=0$).  In this branch, $\gamma(r=0)$ is
 \begin{align}
 \gamma(r=0)=
  \begin{cases}
   \displaystyle {b^2\over a^2+b^2} & (a>b), \\[2ex]
   \displaystyle -{a^2\over a^2+b^2} & (b>a).\\
  \end{cases} 
\label{fsci21Mar21}
\end{align}

\subsection{Various relations}
\label{app:various_rel}

We can always write $\cI_{10}$ and $\cI_{01}$ as follows:
\begin{align}
 x \cI_{10}+y \cI_{01}
 &=
 \int_0^{2\pi}{d\mu\over 2\pi}
 \int_0^{2\pi}{d\nu\over 2\pi}
 {xP\cos\mu +  yQ\cos\nu \over
 1-P\cos\mu - Q\cos\nu}
 \notag\\
 &=
 \int_0^{2\pi}{d\mu\over 2\pi}
 \int_0^{2\pi}{d\nu\over 2\pi}
 \biggl[{x-y\over a^2+b^2}\,
 {b^2 P\cos\mu - a^2  Q\cos\nu \over 1-P\cos\mu - Q\cos\nu}
 \notag\\
 &\hspace{20ex}
 +
 {a^2x+b^2y\over a^2+b^2}\,
 {1-(1-P\cos\mu - Q\cos\nu )
 \over 1-P\cos\mu - Q\cos\nu}\biggr]
 \notag\\
 &={x-y\over a^2+b^2}(b^2 \cI_{10}-a^2 \cI_{01})
 +{a^2x+b^2y\over a^2+b^2}\left({H\over V}+{a^2+b^2\over 4}H-1\right).
\label{iuds16Apr21}
\end{align}

Let us derive the near-ring behavior of $\cI_{00}$.
Near the ring, where $s\to a,w\to b$, we have
\begin{align}
 P\approx {a^2\over a^2+b^2},\quad
 Q\approx {b^2\over a^2+b^2},\qquad
 P+Q\approx 1.
\end{align}
So, in  the integral
\begin{align}
 \cI_{00}=\int{d\mu\over 2\pi}{1\over \sqrt{(1-P\cos\mu)^2-Q^2}},
\end{align}
the dominant contribution comes from $\mu\approx 0$, which corresponds to
the contribution from the part of the ring near the point where we are sitting.  In
this limit, $(1-P\cos\mu)^2-Q^2 \approx (1-P+Q)(D^2+x^2)$, where
$D\equiv \sqrt{1-P-Q}$ is the distance from us to the ring which is now
approximated to be straight, and $x=\sqrt{P/2\,}\,\mu$ is the distance
along the straight ring. So,
\begin{align}
 \cI_{00}
 &\approx
 {1\over 2\pi}\sqrt{2\over (1-P+Q)P}
 \int_{-\Lambda}^{\Lambda} {dx\over\sqrt{D^2+x^2}}
 \notag\\
 &\approx
 {1\over \pi}\sqrt{2\over (1-P+Q)P}\log{\Lambda\over D}
 =
 {1\over \pi}{a^2+b^2\over ab}\log{\Lambda\over D},
\end{align}
where $\Lambda$ is a cutoff which depends on global data and cannot be
determined by a local analysis here.  The relation between $D$ and $u$
works out to be $D\approx {ab\over\sqrt{2}(a^2+b^2)}{1\over u}$.
So, in the end
\begin{align}
 \cI_{00}&\sim {1\over \pi}{a^2+b^2\over ab}\log u.\label{pvw20Apr21}
\end{align}
Evaluation of $\cI_{10}$ is mostly identical, because $\cos\mu\approx
1$ in the $\mu\approx 0$ region. The result is
\begin{align}
 \cI_{10}
 \sim {a^2\over a^2+b^2}
 {1\over \pi}{a^2+b^2\over ab}\log u
 =
 {a\over \pi b}\log u,\qquad
 \cI_{01}
 \sim 
 {b\over \pi a}\log u.\label{pty20Apr21}
\end{align}
This in particular means that
\begin{align}
 \bigtriangleup (b^2 \cI_{10}- a^2 \cI_{01})=\text{regular}\qquad \text{(no $\delta$ function)}.\label{mvhq12Apr21}
\end{align}

Let us briefly discuss how to show
that \eqref{euto20Apr21} is equal to \eqref{eumb20Apr21}, namely
\begin{align}
 b^2\cI_{10}-a^2 \cI_{01}
 &=
 {4ab\,\over(s^2+w^2+a^2+b^2)^2}
 \int_0^{2\pi} {d\mu\over 4\pi}\,
 {(b^2-a^2)sw\cos\mu + ab(s^2-w^2)\over 
 \sqrt{X}(1+\sqrt{X})
 }\,,
\label{ewfj20Apr21}
\end{align}
where $X\equiv 1-(P^2+Q^2)-2PQ\cos\mu$.  Just as we did around
\eqref{jitw19Apr21}, by setting $\nu\to \nu+\mu$ and carrying out the
$\nu$ integral first, we find
\begin{align}
 \cI_{10}=\int_0^{2\pi}{d\mu\over 2\pi}
 {P(P+Q\cos\mu)\over \sqrt{1-S^2}\,(1+\sqrt{1-S^2})}.
\end{align}
$\cI_{01}$ is obtained by $P\leftrightarrow Q$.  Using the relations
\eqref{fsfz13Apr21}, \eqref{fsgo13Apr21}, it is straightforward to
prove~\eqref{ewfj20Apr21}.



\begin{thebibliography}{99}

\bibitem{Behrndt:1997ny} 
  K.~Behrndt, D.~Lust and W.~A.~Sabra,
  ``Stationary solutions of N=2 supergravity,''
  Nucl.\ Phys.\ B {\bf 510}, 264 (1998)
  doi:10.1016/S0550-3213(97)00633-0, 10.1016/S0550-3213(98)81014-6
  [hep-th/9705169].

\bibitem{Gauntlett:2002nw}
J.~P.~Gauntlett, J.~B.~Gutowski, C.~M.~Hull, S.~Pakis and H.~S.~Reall,
``All supersymmetric solutions of minimal supergravity in five- dimensions,''
Class. Quant. Grav. \textbf{20} (2003), 4587-4634
doi:10.1088/0264-9381/20/21/005
[arXiv:hep-th/0209114 [hep-th]].

\bibitem{Bates:2003vx} 
  B.~Bates and F.~Denef,
  ``Exact solutions for supersymmetric stationary black hole composites,''
  JHEP {\bf 1111}, 127 (2011)
  doi:10.1007/JHEP11(2011)127
  [hep-th/0304094].

\bibitem{Bena:2004de}
I.~Bena and N.~P.~Warner,
``One ring to rule them all ... and in the darkness bind them?,''
Adv. Theor. Math. Phys. \textbf{9} (2005) no.5, 667-701
doi:10.4310/ATMP.2005.v9.n5.a1
[arXiv:hep-th/0408106 [hep-th]].

\bibitem{Gauntlett:2004qy} 
  J.~P.~Gauntlett and J.~B.~Gutowski,
  ``General concentric black rings,''
  Phys.\ Rev.\ D {\bf 71}, 045002 (2005)
  doi:10.1103/PhysRevD.71.045002
  [hep-th/0408122].


\bibitem{Bena:2005va} 
  I.~Bena and N.~P.~Warner,
  ``Bubbling supertubes and foaming black holes,''
  Phys.\ Rev.\ D {\bf 74}, 066001 (2006)
  doi:10.1103/PhysRevD.74.066001
  [hep-th/0505166].

\bibitem{Meessen:2006tu} 
  P.~Meessen and T.~Ortin,
  ``The Supersymmetric configurations of N=2, D=4 supergravity coupled to vector supermultiplets,''
  Nucl.\ Phys.\ B {\bf 749}, 291 (2006)
  doi:10.1016/j.nuclphysb.2006.05.025
  [hep-th/0603099].

\bibitem{Ferrara:1995ih} 
  S.~Ferrara, R.~Kallosh and A.~Strominger,
  ``N=2 extremal black holes,''
  Phys.\ Rev.\ D {\bf 52}, R5412 (1995)
  doi:10.1103/PhysRevD.52.R5412
  [hep-th/9508072].

\bibitem{Strominger:1996kf} 
  A.~Strominger,
  ``Macroscopic entropy of N=2 extremal black holes,''
  Phys.\ Lett.\ B {\bf 383}, 39 (1996)
  doi:10.1016/0370-2693(96)00711-3
  [hep-th/9602111].

\bibitem{Ferrara:1996dd} 
  S.~Ferrara and R.~Kallosh,
  ``Supersymmetry and attractors,''
  Phys.\ Rev.\ D {\bf 54}, 1514 (1996)
  doi:10.1103/PhysRevD.54.1514
  [hep-th/9602136].

\bibitem{Ferrara:1996um} 
  S.~Ferrara and R.~Kallosh,
  ``Universality of supersymmetric attractors,''
  Phys.\ Rev.\ D {\bf 54}, 1525 (1996)
  doi:10.1103/PhysRevD.54.1525
  [hep-th/9603090].

\bibitem{Moore:2004fg} 
  G.~W.~Moore,
  ``Strings and Arithmetic,''
  doi:10.1007/978-3-540-30308-4\_8
  hep-th/0401049.

\bibitem{Kraus:2005gh} 
  P.~Kraus and F.~Larsen,
  ``Attractors and black rings,''
  Phys.\ Rev.\ D {\bf 72}, 024010 (2005)
  doi:10.1103/PhysRevD.72.024010
  [hep-th/0503219].

\bibitem{Larsen:2006xm} 
  F.~Larsen,
  ``The Attractor Mechanism in Five Dimensions,''
  Lect.\ Notes Phys.\  {\bf 755}, 249 (2008)
  [hep-th/0608191].

\bibitem{Denef:2000ar} 
  F.~Denef,
  ``On the correspondence between D-branes and stationary supergravity solutions of type II Calabi-Yau compactifications,''
  hep-th/0010222.

\bibitem{Denef:2001ix} 
  F.~Denef,
  ``(Dis)assembling special Lagrangians,''
  hep-th/0107152.

\bibitem{Moore2010pitp}
G.~W. Moore, ``{PiTP lectures on BPS states and wall-crossing in d= 4, N= 2
  theories},''.
  \url{http://www.physics.rutgers.edu/~gmoore/PiTP_July26_2010.pdf}.

\bibitem{Denef:2007vg} 
  F.~Denef and G.~W.~Moore,
  ``Split states, entropy enigmas, holes and halos,''
  JHEP {\bf 1111}, 129 (2011)
  doi:10.1007/JHEP11(2011)129
  [hep-th/0702146].


\bibitem{Berglund:2005vb} 
  P.~Berglund, E.~G.~Gimon and T.~S.~Levi,
  ``Supergravity microstates for BPS black holes and black rings,''
  JHEP {\bf 0606}, 007 (2006)
  doi:10.1088/1126-6708/2006/06/007
  [hep-th/0505167].

\bibitem{Heidmann:2018vky} 
  P.~Heidmann and S.~Mondal,
  ``The full space of BPS multicenter states with pure D-brane charges,''
  JHEP {\bf 1906}, 011 (2019)
  doi:10.1007/JHEP06(2019)011
  [arXiv:1810.10019 [hep-th]].

\bibitem{Mateos:2001qs} 
  D.~Mateos and P.~K.~Townsend,
  ``Supertubes,''
  Phys.\ Rev.\ Lett.\  {\bf 87}, 011602 (2001)
  doi:10.1103/PhysRevLett.87.011602
  [hep-th/0103030].

\bibitem{Park:2015gka} 
  M.~Park and M.~Shigemori,
  ``Codimension-2 solutions in five-dimensional supergravity,''
  JHEP {\bf 1510}, 011 (2015)
  doi:10.1007/JHEP10(2015)011
  [arXiv:1505.05169 [hep-th]].

\bibitem{Fernandez-Melgarejo:2017dme} 
  J.~J.~Fernandez-Melgarejo, M.~Park and M.~Shigemori,
  ``Non-Abelian Supertubes,''
  JHEP {\bf 1712}, 103 (2017)
  doi:10.1007/JHEP12(2017)103
  [arXiv:1709.02388 [hep-th]].

\bibitem{Lunin:2001jy} 
  O.~Lunin and S.~D.~Mathur,
  ``AdS / CFT duality and the black hole information paradox,''
  Nucl.\ Phys.\ B {\bf 623}, 342 (2002)
  doi:10.1016/S0550-3213(01)00620-4
  [hep-th/0109154].

\bibitem{Lunin:2002iz} 
  O.~Lunin, J.~M.~Maldacena and L.~Maoz,
  ``Gravity solutions for the D1-D5 system with angular momentum,''
  hep-th/0212210.

\bibitem{Niehoff:2013kia} 
  B.~E.~Niehoff and N.~P.~Warner,
  ``Doubly-Fluctuating BPS Solutions in Six Dimensions,''
  JHEP {\bf 1310}, 137 (2013)
  doi:10.1007/JHEP10(2013)137
  [arXiv:1303.5449 [hep-th]].

\bibitem{Bena:2005ni} 
  I.~Bena, P.~Kraus and N.~P.~Warner,
  ``Black rings in Taub-NUT,''
  Phys.\ Rev.\ D {\bf 72}, 084019 (2005)
  doi:10.1103/PhysRevD.72.084019
  [hep-th/0504142].

\bibitem{Gregory:1997te} 
  R.~Gregory, J.~A.~Harvey and G.~W.~Moore,
  ``Unwinding strings and t duality of Kaluza-Klein and h monopoles,''
  Adv.\ Theor.\ Math.\ Phys.\  {\bf 1}, 283 (1997)
  doi:10.4310/ATMP.1997.v1.n2.a6
  [hep-th/9708086].

\bibitem{Marolf:2000cb}
D.~Marolf,
``Chern-Simons terms and the three notions of charge,''
[arXiv:hep-th/0006117 [hep-th]].

\bibitem{Schwarz:1982ec}
A.~S.~Schwarz,
``FIELD THEORIES WITH NO LOCAL CONSERVATION OF THE ELECTRIC CHARGE,''
Nucl. Phys. B \textbf{208} (1982), 141-158
doi:10.1016/0550-3213(82)90190-0

\bibitem{Gutowski:2004yv}
J.~B.~Gutowski and H.~S.~Reall,
``General supersymmetric AdS(5) black holes,''
JHEP \textbf{04} (2004), 048
doi:10.1088/1126-6708/2004/04/048
[arXiv:hep-th/0401129 [hep-th]].

\bibitem{Gutowski:2005id}
J.~B.~Gutowski and W.~Sabra,
``General supersymmetric solutions of five-dimensional supergravity,''
JHEP \textbf{10} (2005), 039
doi:10.1088/1126-6708/2005/10/039
[arXiv:hep-th/0505185 [hep-th]].

\bibitem{Hanaki:2007mb} 
  K.~Hanaki, K.~Ohashi and Y.~Tachikawa,
  ``Comments on charges and near-horizon data of black rings,''
  JHEP {\bf 0712}, 057 (2007)
  doi:10.1088/1126-6708/2007/12/057
  [arXiv:0704.1819 [hep-th]].

\bibitem{Denef:2000nb} 
  F.~Denef,
  ``Supergravity flows and D-brane stability,''
  JHEP {\bf 0008}, 050 (2000)
  doi:10.1088/1126-6708/2000/08/050
  [hep-th/0005049].

\bibitem{DallAgata:2010srl} 
  G.~Dall'Agata, S.~Giusto and C.~Ruef,
  ``U-duality and non-BPS solutions,''
  JHEP {\bf 1102}, 074 (2011)
  doi:10.1007/JHEP02(2011)074
  [arXiv:1012.4803 [hep-th]].

\bibitem{Strominger:1996sh} 
  A.~Strominger and C.~Vafa,
  ``Microscopic origin of the Bekenstein-Hawking entropy,''
  Phys.\ Lett.\ B {\bf 379}, 99 (1996)
  doi:10.1016/0370-2693(96)00345-0
  [hep-th/9601029].

\bibitem{Rychkov:2005ji} 
  V.~S.~Rychkov,
  ``D1-D5 black hole microstate counting from supergravity,''
  JHEP {\bf 0601}, 063 (2006)
  doi:10.1088/1126-6708/2006/01/063
  [hep-th/0512053].

\bibitem{Krishnan:2015vha} 
  C.~Krishnan and A.~Raju,
  ``A Note on D1-D5 Entropy and Geometric Quantization,''
  JHEP {\bf 1506}, 054 (2015)
  doi:10.1007/JHEP06(2015)054
  [arXiv:1504.04330 [hep-th]].

\bibitem{Kanitscheider:2007wq} 
  I.~Kanitscheider, K.~Skenderis and M.~Taylor,
  ``Fuzzballs with internal excitations,''
  JHEP {\bf 0706}, 056 (2007)
  doi:10.1088/1126-6708/2007/06/056
  [arXiv:0704.0690 [hep-th]].

\bibitem{Giusto:2013bda} 
  S.~Giusto and R.~Russo,
  ``Superdescendants of the D1D5 CFT and their dual 3-charge geometries,''
  JHEP {\bf 1403}, 007 (2014)
  doi:10.1007/JHEP03(2014)007
  [arXiv:1311.5536 [hep-th]].

\bibitem{Bena:2008dw} 
  I.~Bena, N.~Bobev, C.~Ruef and N.~P.~Warner,
  ``Supertubes in Bubbling Backgrounds: Born-Infeld Meets Supergravity,''
  JHEP {\bf 0907}, 106 (2009)
  doi:10.1088/1126-6708/2009/07/106
  [arXiv:0812.2942 [hep-th]].

\bibitem{Giusto:2019qig} 
  S.~Giusto, S.~Rawash and D.~Turton,
  ``Ads$_{3}$ holography at dimension two,''
  JHEP {\bf 1907}, 171 (2019)
  doi:10.1007/JHEP07(2019)171
  [arXiv:1904.12880 [hep-th]].

\bibitem{deBoer:2012ma} 
  J.~de Boer and M.~Shigemori,
  ``Exotic Branes in String Theory,''
  Phys.\ Rept.\  {\bf 532}, 65 (2013)
  doi:10.1016/j.physrep.2013.07.003
  [arXiv:1209.6056 [hep-th]].

\bibitem{Peet:2000hn} 
  A.~W.~Peet,
  ``TASI lectures on black holes in string theory,''
  doi:10.1142/9789812799630\_0003
  hep-th/0008241.


\bibitem{Emparan:2001ux} 
  R.~Emparan, D.~Mateos and P.~K.~Townsend,
  ``Supergravity supertubes,''
  JHEP {\bf 0107}, 011 (2001)
  doi:10.1088/1126-6708/2001/07/011
  [hep-th/0106012].

\bibitem{Kanitscheider:2006zf} 
  I.~Kanitscheider, K.~Skenderis and M.~Taylor,
  ``Holographic anatomy of fuzzballs,''
  JHEP {\bf 0704}, 023 (2007)
  doi:10.1088/1126-6708/2007/04/023
  [hep-th/0611171].

\bibitem{Skenderis:2006ah} 
  K.~Skenderis and M.~Taylor,
  ``Fuzzball solutions and D1-D5 microstates,''
  Phys.\ Rev.\ Lett.\  {\bf 98}, 071601 (2007)
  doi:10.1103/PhysRevLett.98.071601
  [hep-th/0609154].

\bibitem{David:2002wn} 
  J.~R.~David, G.~Mandal and S.~R.~Wadia,
  ``Microscopic formulation of black holes in string theory,''
  Phys.\ Rept.\  {\bf 369}, 549 (2002)
  doi:10.1016/S0370-1573(02)00271-5
  [hep-th/0203048].

\bibitem{Avery:2010qw} 
  S.~G.~Avery,
  ``Using the D1D5 CFT to Understand Black Holes,''
  arXiv:1012.0072 [hep-th].

\bibitem{Bena:2017xbt} 
  I.~Bena, S.~Giusto, E.~J.~Martinec, R.~Russo, M.~Shigemori, D.~Turton and N.~P.~Warner,
  ``Asymptotically-flat supergravity solutions deep inside the black-hole regime,''
  JHEP {\bf 1802}, 014 (2018)
  doi:10.1007/JHEP02(2018)014
  [arXiv:1711.10474 [hep-th]].

\bibitem{Schwimmer:1986mf}
A.~Schwimmer and N.~Seiberg,
``Comments on the N=2, N=3, N=4 Superconformal Algebras in Two-Dimensions,''
Phys. Lett. B \textbf{184} (1987), 191-196
doi:10.1016/0370-2693(87)90566-1

\bibitem{Bena:2008wt} 
  I.~Bena, N.~Bobev and N.~P.~Warner,
  ``Spectral Flow, and the Spectrum of Multi-Center Solutions,''
  Phys.\ Rev.\ D {\bf 77}, 125025 (2008)
  doi:10.1103/PhysRevD.77.125025
  [arXiv:0803.1203 [hep-th]].

\bibitem{Giusto:2012yz} 
  S.~Giusto, O.~Lunin, S.~D.~Mathur and D.~Turton,
  ``D1-D5-P microstates at the cap,''
  JHEP {\bf 1302}, 050 (2013)
  doi:10.1007/JHEP02(2013)050
  [arXiv:1211.0306 [hep-th]].

\bibitem{Hampton:2018ygz} 
  S.~Hampton, S.~D.~Mathur and I.~G.~Zadeh,
  ``Lifting of D1-D5-P states,''
  JHEP {\bf 1901}, 075 (2019)
  doi:10.1007/JHEP01(2019)075
  [arXiv:1804.10097 [hep-th]].

\bibitem{Lunin:2002fw} 
  O.~Lunin and S.~D.~Mathur,
  ``Rotating deformations of AdS(3) x S**3, the orbifold CFT and strings in the pp wave limit,''
  Nucl.\ Phys.\ B {\bf 642}, 91 (2002)
  doi:10.1016/S0550-3213(02)00677-6
  [hep-th/0206107].

\bibitem{Gomis:2002qi} 
  J.~Gomis, L.~Motl and A.~Strominger,
  ``PP wave / CFT(2) duality,''
  JHEP {\bf 0211}, 016 (2002)
  doi:10.1088/1126-6708/2002/11/016
  [hep-th/0206166].


\bibitem{Gava:2002xb} 
  E.~Gava and K.~S.~Narain,
  ``Proving the PP wave / CFT(2) duality,''
  JHEP {\bf 0212}, 023 (2002)
  doi:10.1088/1126-6708/2002/12/023
  [hep-th/0208081].

\bibitem{Bena:2011zw} 
  I.~Bena, B.~D.~Chowdhury, J.~de Boer, S.~El-Showk and M.~Shigemori,
  ``Moulting Black Holes,''
  JHEP {\bf 1203}, 094 (2012)
  doi:10.1007/JHEP03(2012)094
  [arXiv:1108.0411 [hep-th]].

\bibitem{GR}
I.~S.~Gradshteyn and, I.~M.~Ryzhik,
``Table of Integrals, Series and Products''.

\bibitem{functions.wolfram}
http://functions.wolfram.com/08.02.17.0003.01
\end{thebibliography}
\end{document}